\documentclass[fleqn,usenatbib]{mnras}

\usepackage{newtxtext,newtxmath}
\usepackage[mathscr]{euscript}
\usepackage[T1]{fontenc}
\usepackage{ae,aecompl}
\usepackage[utf8]{inputenc}

\usepackage{graphicx}	% Including figure files
\usepackage{amsmath}	% Advanced maths commands
\usepackage{amssymb}	% Extra maths symbols
\usepackage{bm}
\usepackage{cancel} 
\usepackage{color}
\definecolor{azure}{rgb}{0.0, 0.5, 1.0}
\usepackage{graphicx}	% Including figure files
\usepackage{amssymb}
\usepackage{bm,tensor,braket}
\usepackage{xcolor}
\usepackage{enumerate}
\usepackage{mathtools}
\usepackage[normalem]{ulem}
\newcommand{\rmd}{{\rm d}}

\newcommand{\Reply}[1]{{#1}}
\title[Triaxially-deformed Freely-precessing NSs]{Triaxially-deformed
Freely-precessing Neutron Stars: Continuous electromagnetic and
gravitational radiation}

\author[Y. Gao et al.]{
Yong Gao,$^{1,2}$
Lijing Shao,$^{2,3}$\thanks{Corresponding author. E-mail: lshao@pku.edu.cn (LS)}
Rui Xu,$^{2}$
Ling Sun,$^{4,5}$
Chang Liu,$^{1,2}$
and Ren-Xin Xu$^{1,2}$
\\
$^{1}$Department of Astronomy, School of Physics, Peking University,
Beijing 100871, China\\
$^{2}$Kavli Institute for Astronomy and Astrophysics, Peking University,
Beijing 100871, China\\
$^{3}$National Astronomical Observatories, Chinese Academy of Sciences,
Beijing 100012, China \\
$^{4}$LIGO Laboratory, California Institute of Technology, Pasadena,
California 91125, USA \\
$^{5}$Centre for Gravitational Astrophysics, College of Science, The Australian National University, ACT 2601, Australia
}

\date{Accepted XXX. Received YYY; in original form ZZZ}

\pubyear{2020}

\begin{document}
\label{firstpage}
\pagerange{\pageref{firstpage}--\pageref{lastpage}}
\maketitle

\begin{abstract}
The shape of a neutron star (NS) is closely linked to its internal
structure and the equation of state of supranuclear matters. A rapidly
rotating, asymmetric NS in the Milky Way undergoes free precession, making it a
potential source for {\it multimessenger} observation.
The free precession could manifest in (i) the spectra of continuous
gravitational waves (GWs) in the kilohertz band for ground-based GW
detectors, and (ii) the timing behavior and pulse-profile
characteristics if the NS is monitored as a pulsar with radio and/or
X-ray telescopes. We extend previous work and investigate in great
detail the free precession of a triaxially deformed NS with analytical
and numerical approaches. In particular, its associated continuous GWs
and pulse signals are derived. Explicit examples are illustrated for
the continuous GWs, as well as timing residuals in both time and
frequency domains. These results are ready to be used for future
multimessenger observation of triaxially-deformed freely-precessing
NSs, in order to extract scientific implication as much as possible.
\end{abstract}


%%%%%%%%%%%%%%%%%%%%%%%%%%%%%%%%%%%%%%%%%%%%%%%%%%

%%%%%%%%%%%%%%%%% BODY OF PAPER %%%%%%%%%%%%%%%%%%

%---------------------------------------------------------------------
\section{Introduction}
\label{sec:intro}

Pulsars are magnetized rotating neutron stars (NSs). Using the so-called
pulsar timing technique, the Hulse-Taylor pulsar provided the first
validation for the existence of gravitational waves
\citep[GWs;][]{Taylor:1979zz}. In the new era after the direct observation
of GWs with ground-based laser interferometric detectors
\citep{TheLIGOScientific:2014jea,TheVirgo:2014hva,Abbott:2016blz, TheLIGOScientific:2017qsa, LIGOScientific:2018mvr},
pulsars continue to play an important role in the context of GW
astrophysics. They can be perceived as GW {\it sources} radiating
continuous GWs, as well as GW {\it detectors} in the form of pulsar timing
arrays \citep{Janssen:2014dka, Perera:2019sca}. In this work, we are
interested in freely precessing, asymmetric NSs, which can produce both
modulated pulse signals and continuous GW radiations
with characteristic features \citep{Zimmermann:1979ip,Zimmermann:1980ba, Bisnovatyi-Kogan:1990, 
Jones:2000ud, Jones:2001yg}, and thus become potential multimessenger
sources of great scientific interest.
%and provide maximal scientific payoff from this class of astrophysical objects.

The most compelling evidence for NS free precession comes from
PSR~B1828$-$11. Timing observation over 13 years for this isolated pulsar
showed strong Fourier power at periods of about 250, 500 and 1000 days
\citep{Stairs:2000zz}, which could be an indication for free precession.
\citet{Link:2001zr} suggested the period at 500 days as the precession
period of a biaxial NS with a precessing angle of $\sim 3^{\circ}$ and a
dipole moment nearly orthogonal to the symmetric axis. They also
interpreted the period at 250 days as a result of the electromagnetic
dipole torque. Timing data of another pulsar, PSR~B1642$-$03, provided
additional support for the idea of NS free precession
\citep{1993ASPC...36...43C, Shabanova:2001ud}. \Reply{Recently, the
CHIME/FRB Collaboration reported the detection of a $16.35\pm0.15$ day
periodicity in the radio burst from FRB~180916.J0158$+$65
\citep{Amiri:2020gno}. This periodicity may arise from the free precession
of a magnetar \citep{Zanazzi:2020vyp, Levin:2020rhj}. Another possible
evidence of precession comes from oscillations in gamma-ray burst
afterglows. The afterglow may be powered by a long-lived remnant
\citep{Dai:1998hm, Zhang:2005fa}, which might be a millisecond magnetar
\citep{Suvorov:2020eji}. In the early stages of its life, the magnetar is
likely to precess and result in modulation of X-ray luminosity.
\citet{Suvorov:2020eji} found that the data of two bursts are highly
consistent with precessing oblique rotators.} While more evidence is needed
to solidify the phenomenon of NS free precession, it is nowadays certainly
interesting to study it in the context of GW astrophysics.

The conventional model for NS structure consists of a liquid core and a
thin solid crust. \citet{Jones:2000ud} conjectured that the core of a NS
does not participate in the free precession. Assuming the crust-only
precession, they constructed a simple model with a thin radio beam fixed on
the body of a biaxially deformed NS, and they assumed that the beam is
aligned with the dipole moment. The model was applied to some potential
candidates which may be undergoing free precession \citep{Jones:2000ud}.
Although our understanding related to NSs has advanced
remarkably~\citep{2018RPPh...81e6902B}, NS structure is still unclear and
alternative models have already been proposed. NSs could actually be
strange stars if Witten's conjecture is correct~\citep{Witten:1984rs}, and
they could globally be in a solid state if quarks are condensed in position
space~\citep{Xu:2003xe} or momentum space~\citep{Mannarelli:2007bs},
resulting in highly elastic quadrupole deformations~\citep{Owen:2005fn} and
thus large free precession amplitudes. Therefore, a multimessenger study of
freely-precessing NSs would help in understating the equation of state of
cold matter at supranuclear density and distinguish between different
models.

Previous studies dominantly focused on biaxial NSs. In the most generic
case, the deformation of a NS does not need to be biaxial. A triaxially
deformed NS can demonstrate new features in its free precession. We extend
the simple model in \citet{Jones:2000ud} and study the timing residual of a
freely precessing triaxial NS in this work. The internal dissipation from
the frictional-type coupling between the crust and the core may damp the
wobble angle in a relatively short timescale \citep{Jones:2001yg}. As an
illustrative work, we do not consider the damping here, but use different
wobble angles in our calculation, from large ones to small ones, to display
the modulations of spin period and spin period derivative in both time and
frequency domains. The precession modulates the pulse width as well, which
provides a good way to probe the beam shape of the pulsar radiation
\citep{Link:2001zr, Desvignes:2019uxs}. We use a simple cone model
\citep{Gil:1984ads, Lorimer:2005misc} to study the pulse-width modulation
of triaxially-deformed freely-precessing NSs, and investigate the change of
pulse width with different choices of wobble angles.

From the GW perspectives, precessing NSs have been recognized as
potential sources of continuous GWs for decades \citep{Zimmermann:1979ip,
Zimmermann:1980ba, Alpar:1985kz}. In the new era of GW astronomy, the
detection of GWs from precessing NSs with ground-based detectors is
imminent. Using the Advanced LIGO data from its first and second observing
runs, the search of continuous GWs at once and twice rotation frequencies
from 222 pulsars has been performed \citep{Authors:2019ztc}. Stringent
upper limits are set on the GW amplitude, the fiducial ellipticity, and the
mass quadrupole moment via the search at the twice of the rotation
frequency. These results can be used for testing various alternatives to
the General Relativity \citep[e.g.,][]{Xu:2019gua}.

\citet{Zimmermann:1980ba} treated precessing triaxial NSs as rigid bodies
and derived the quadrupole waveform for them. In addition, he simplified
the waveform assuming a small wobble angle, and showed that the spectral
lines of the continuous GWs are located at angular frequencies of
$\Omega_{\textrm{r}}+\Omega_{\textrm{p}}$ and $2\Omega_{\textrm{r}}$, where
$\Omega_{\textrm{r}}$ is the rotation angular frequency, and
$\Omega_{\textrm{p}}$ is the free precession angular frequency of the NS.
The first-order spectral lines yield little information about other
physical properties of NSs beyond the rotation and precession frequencies.
Based on \citet{Zimmermann:1980ba}, \citet{VanDenBroeck:2004wj} obtained a
third angular frequency at
$2\left(\Omega_{\textrm{r}}+\Omega_{\textrm{p}}\right)$ by expanding the
waveform to the second order of the wobble angle. The feasibility to detect
continuous GWs from precessing triaxial NSs was re-examined, and it is
found that the deviation from axisymmetry, the oblateness, and the wobble
angle can be determined if the second-order line is observed
\citep{VanDenBroeck:2004wj}.

Following \citet{Zimmermann:1980ba} and \citet{VanDenBroeck:2004wj}, in
this work we use a Newtonian treatment for the precession, augmented with
the GW radiation formalism in GR \citep{Misner:1974qy}. In this problem, the
Newtonian treatment is indeed also valid for strong-field objects like NSs,
if the GR expressions for the integrals of various moments are used
\citep{Thorne:1980ru}. Similarly to \citet{VanDenBroeck:2004wj}, we expand
the GW waveform to the second order of the precessing angle. However,
unlike \citet{VanDenBroeck:2004wj}, where a hierarchy for small
parameters is assumed, more generically we treat the deviation from axisymmetry as a
small parameter independent of the wobble angle in the expansion.
Consequently, we obtain three more frequencies in the continuous GW spectra
that are useful for a more complete extraction of physical information.

The structure of this paper is as follows. In Section \ref{sec:free_prec},
we provide both analytical and numerical solutions for freely precessing
triaxial rigid bodies. Estimations of the oblateness, the nonaxisymmetry,
and the wobble angle for elastically deformed NSs are given based on
existing literature. In Section \ref{sec:pulsar_radiation}, we show the
timing residuals and pulse-width modulations of precessing triaxial NSs.
These features could be identified if the NS is observed as a radio and/or
X-ray pulsar. In Section \ref{sec:gw}, after briefly reviewing the
quadrupole formula in \citet{Zimmermann:1980ba}, we expand the waveform to
the second order assuming a small wobble angle, a small nonaxisymmetry and
a small oblateness. Because of the relaxation in the assumption about the
small quantities, three new spectral lines are obtained with respect to
previous studies. In Section \ref{sec:disc}, we discuss the extraction of
physical information of NSs from radio signals and continuous GWs. We
briefly summarize our work in Section \ref{sec:sum}.

%---------------------------------------------------------------------
\section{Free precession of triaxial rigid bodies} 
\label{sec:free_prec}
%--------------------------------------------------------------------- 

\begin{figure}
    \centering
    \includegraphics[width=8.2cm]{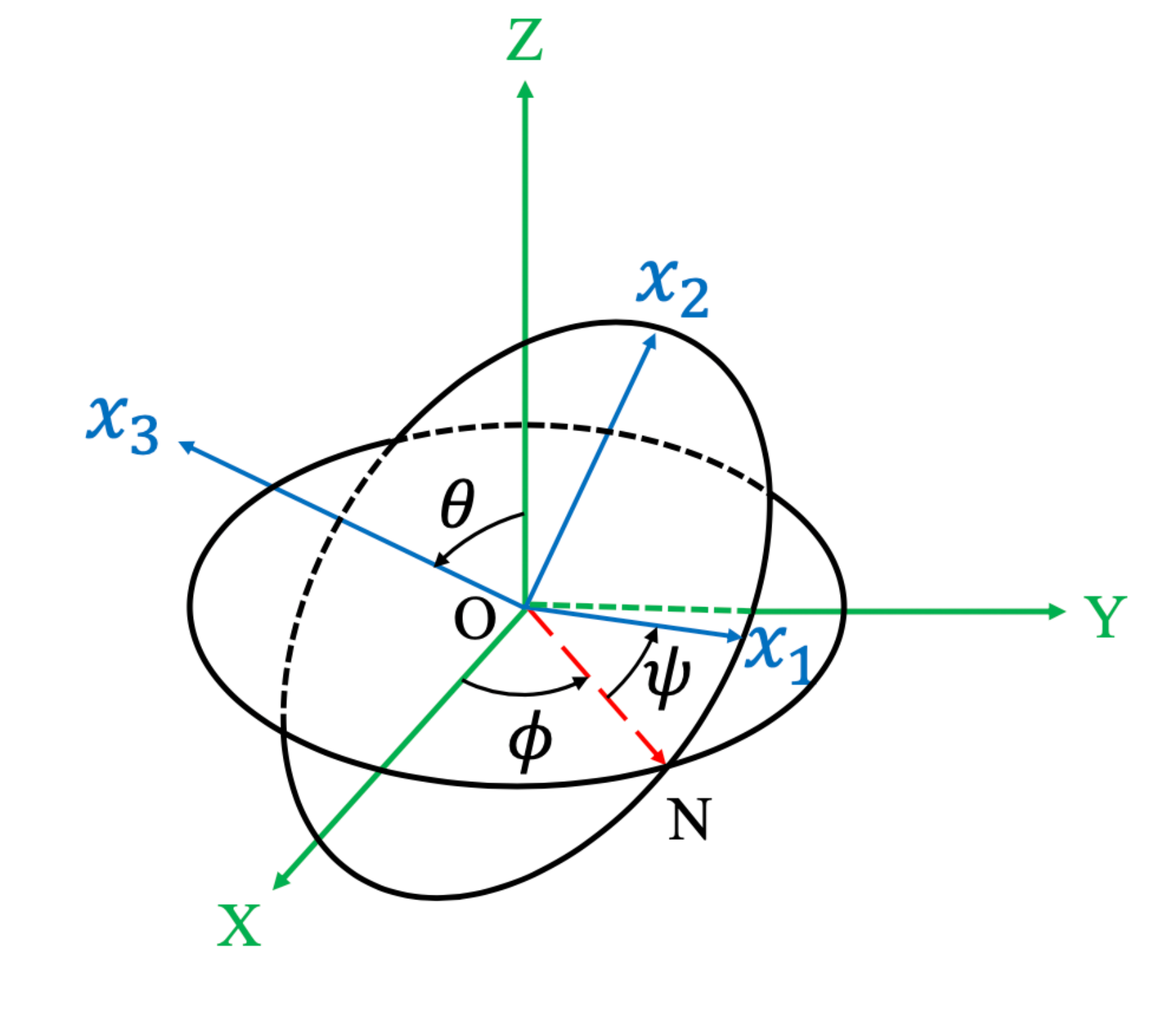}
    \caption{The inertial and body coordinate systems for the rigid body.
    Uppercase letters, $\textrm{X}$, $\textrm{Y}$, and $\textrm{Z}$, denote
    the inertial frame coordinates, while lowercase letters, $x_{1}$,
    $x_{2}$, and $x_{3}$, denote the coordinates in the body frame. Three
    Euler angles, $\theta$, $\phi$, and $\psi$, are defined as shown. }
    \label{fig.euler_angles}
\end{figure}

In general, the rotation axis of a rigid body does not coincide with its
principal axes. As a consequence, a freely rotating rigid body precesses
around the direction of the total angular momentum
\citep{landau1960course}. The motion of the body can be described by three
Euler angles, $\theta$, $\phi$, and $\psi$, and their time derivatives. In
Fig.~\ref{fig.euler_angles} we denote the coordinates of the inertial
reference frame by uppercase letters, $\textrm{X}$, $\textrm{Y}$, and
$\textrm{Z}$, with unit basis vectors $\widehat{e}_{\textrm{X}}$,
$\widehat{e}_{\textrm{Y}}$, and $\widehat{e}_{\textrm{Z}}$. The vector
$\widehat{e}_{\textrm{Z}}$ is chosen to be in the direction of the angular
momentum of the rigid body, $\mathbf{J}$. We use lowercase letters,
$x_{1}$, $x_{2}$, and $x_{3}$, to denote the coordinates in the body frame, which is attached
on the rigid body. Their unit basis vectors are $\widehat{e}_{1}$,
$\widehat{e}_{2}$, and $\widehat{e}_{3}$, chosen to be parallel to the three
individual eigenvectors of the moment of inertia tensor. We use $I_{1}$,
$I_{2}$, and $I_{3}$ as the diagonal components of the moment of inertia
tensor in the body frame.

%--
%--

For freely precessing rigid bodies, the dynamical equations of motion in
the body frame are \citep{landau1960course},
%--
\begin{align}
    I_1 \dot {\omega}_1 - \left( I_2 - I_3 \right) %\cdot
    \omega_2 \omega_3 &= 0 \label{eqn:euler_body1}\,, \\
    I_2 \dot {\omega}_2 - \left( I_3 - I_1 \right) %\cdot
    \omega_3 \omega_1 &= 0 \label{eqn:euler_body2}\,, \\
    I_3 \dot {\omega}_3 - \left( I_1 - I_2 \right) %\cdot
    \omega_1 \omega_2 &= 0 \label{eqn:euler_body3}\,,
\end{align}
%--
where $\omega_1$, $\omega_2$, and $\omega_3$ represent the angular
velocities along $\widehat{e}_{1}$, $\widehat{e}_{2}$, and
$\widehat{e}_{3}$. The dots denote the derivatives with respect to time $t$.
Considering the kinematics of the rigid body, the evolution of the three
Euler angles can be described by \citep{landau1960course},
%--
\begin{align}
    \omega_{1} &={\dot \phi} \sin \theta \sin \psi +{\dot \theta}\cos \psi
    \label{eqn:kinetic1} \,, \\
    \omega_{2} &={\dot \phi} \sin \theta \cos \psi-{\dot \theta}\sin \psi
    \label{eqn:kinetic2}\,, \\
    \omega_{3} &={\dot \phi} \cos \theta+ \dot \psi \label{eqn:kinetic3}\,,
\end{align}
%--
where the Euler angles are defined in Fig.~\ref{fig.euler_angles}. 

As the rigid body is torque free, both the kinetic energy
%--
\begin{equation}
    E=\frac{1}{2}\left(I_{1} \omega_{1}^{2}+I_{2} \omega_{2}^{2}+I_{3}
    \omega_{3}^{2}\right)\,,
\end{equation} 
%--
and the angular momentum
%--
\begin{equation}
    J =\left(I_{1}^{2} \omega_{1}^{2}+I_{2}^{2} \omega_{2}^{2}+I_{3}^{2}
    \omega_{3}^{2}\right)^{1 /2}\,,
\end{equation}
%--
are conserved. In the following, we assume that the principal moments of
inertia satisfy $I_{1} < I_{2} < I_{3}$. We also assume $J^{2}>2 E I_{2}$,
which is equivalent to the condition that the tail of the angular momentum
$\mathbf{J}$ moves around $\widehat{e}_{3}$ along a closed curve in the
body frame \citep{landau1960course}. Results for other choices can be
obtained by properly relabeling the indices.

The motion of a rigid body described by
Eqs.~(\ref{eqn:euler_body1}--\ref{eqn:kinetic3}) is an initial value
problem. In principle, one can obtain the evolution of the orientation of
the triaxial rigid body at any time once the initial values of the three
Euler angles and the angular velocities are specified. In subsequent
calculations, we choose the initial values such that at $t=0$, one has
$\phi = 0$, $\psi=\pi/2$, and $\theta$ is at its minimum value $\theta_{\rm
min}$. The initial values of the angular velocities in the body frame are
denoted as $\omega_1=a$, $\omega_2=0$, and $\omega_3=b$ at $t=0$. These
assumptions can easily be extended to generic cases. \Reply{Worth to
stress that, throughout the calculation, we have assumed that the moments
$I_{i}$ ($i=1, 2, 3$) are constant and the body undergoes free precession.
We have ignored the damping of the precession due to some physical
processes for NSs (e.g., the fluid dynamics of the NS interior).} Now, we
discuss the analytical solution and the numerical method to solve the
equations of motion in Eqs.~(\ref{eqn:euler_body1}--\ref{eqn:kinetic3}).

%---------------------------------------------------------------------
\subsection{Analytical solution}
\label{sec:analy_solution}
%---------------------------------------------------------------------

The exact analytical solution to
Eqs.~(\ref{eqn:euler_body1}--\ref{eqn:kinetic3}) for a precessing triaxial
rigid body has been obtained in terms of the elliptic functions
\citep{landau1960course, Zimmermann:1980ba, whittaker1988treatise, Shakura:1998ps, 
VanDenBroeck:2004wj, Akgun:2005nd, Pina2015DrawingTF, Lasky:2013bpa}. Here we briefly review
the solution according to \citet{landau1960course} for readers'
convenience.

The angular velocities in the body frame are,
%--
\begin{align}
    &\omega_{1}(\tau)= a\, \mathtt{cn} (\tau, m) \label{eqn:omega1}\,,\\
    &\omega_{2}(\tau)=a\left[\frac{I_{1}\left(I_{3}-I_{1}\right)}{I_{2}
    \left(I_{3} - I_{2}\right)} \right]^{1 / 2} \mathtt{sn} (\tau,
    m)\,,\\
    &\omega_{3}(\tau)=b \, \mathtt{dn}(\tau, m)\,,
\end{align}
%--
where $\tau$ is the dimensionless time variable,
%--
\begin{equation}
    \tau =t \sqrt{\frac{\left(I_{3}-I_{2}\right)\left(J^{2}-2 E
    I_{1}\right)}{I_{1} I_{2} I_{3}}} \,,
\end{equation}
%--
and $\mathtt{sn}$, $\mathtt{cn}$, and $\mathtt{dn}$ are the elliptic
functions \citep[see e.g.,][]{olver2010nist}. The parameter $m$ can be expressed as,
%--
\begin{equation}
\label{eqn:modulus}
    m =\frac{\left(I_{2}-I_{1}\right) I_{1} a^{2}}{\left(I_{3}-I_{2}\right)
    I_{3} b^{2}}\,.
\end{equation}
%--

The angular velocities in the body frame are periodic with a period,
%--
\begin{equation}
    T=\frac{4 K(m)}{b}\left[\frac{I_{1}
    I_{2}}{\left(I_{3}-I_{1}\right)\left(I_{3}-I_{2}\right)}\right]^{1 /
    2}\,,
\end{equation}
%--
where $K(m)$ is the complete elliptic integral of the first kind
\citep{olver2010nist}. The period $T$ is the free precession period. 
If $I_{1}$ is nearly equal to $I_{2}$, the
parameter $m$ is close to zero. In this case, the period $T$ is
approximately $2 \pi I_{1} /\left[\omega_{3}\left(I_{3} -
I_{1}\right)\right]$, which is the well-known free precession period for a
biaxial body.

The Euler angles $\theta$ and $\psi$ are also periodic, and
can be expressed as,
%--
\begin{align}
    &\cos \theta=\frac{I_{3} b}{J} \mathtt{dn} (\tau,m)\,,\\
    &\tan \psi = \left[ \frac{I_{1}\left( I_{3}-I_{2}\right)}{I_{2}
    \left(I_{3}-I_{1}\right)}\right]^{1 / 2} \frac{\mathtt{cn}(\tau,
    m)}{\mathtt{sn}(\tau, m)}\,.
\end{align}
%--
From the above two equations, one finds that the angle $\theta$ has a period
of $T/2$, while the angle $\psi$ has a period of $T$. In contrast, the
angle $\phi$ is not periodic. It can be represented as a sum of two parts,
$\phi = \phi_{1} + \phi_{2}$. The ``periodic'' part $\phi_{1}$ has a period
of $T/2$, and is defined via,
%--
\begin{equation}
\label{eq:phi1}
    \exp \left[2 {\rm i}
    \phi_{1}(t)\right]=\frac{\vartheta_{4}\left(\frac{2 \pi t}{T}+ {\rm i}
    \pi \alpha, q \right)}{\vartheta_{4}\left(\frac{2 \pi t}{T}- {\rm i}
    \pi \alpha ,q \right)}\,,
\end{equation}
%--
where $\vartheta_{4}$ is the fourth Jacobi theta function with nome \mbox{$q=\exp
\left[-\pi K(1-m)/K(m) \right]$}. In Eq.~(\ref{eq:phi1}), $\alpha$ is
determined via
%--
\begin{equation}
    \mathtt{sn} \left[2 \mathrm{i} \alpha K(m) \right]=\frac{\mathrm{i}
    I_{3} b}{I_{1} a}\,.
\end{equation}
%--
The ``linear-in-time'' part $\phi_{2}$ is given by,
%--
\begin{equation}
\label{eq:phi2}
    {\phi_{2}=\frac{2 \pi t}{T_{1}}} = \left( \frac{J}{I_{1}}+\frac{2\pi
    \mathrm{i}}{T} \frac{\vartheta_{4}^{\prime}(\mathrm{i} \pi \alpha,
    q)}{\vartheta_{4}(\mathrm{i} \pi \alpha, q)} \right) t \,,
\end{equation}
%--
where $\vartheta_{4}^{\prime}(u,q)$ is the derivative of
$\vartheta_{4}(u,q)$ with respect to $u$.\footnote{Note that the solution
of the Euler angle $\phi$ in \citet{landau1960course} and
\citet{Zimmermann:1980ba} has sign typos when the following theta function
and its derivative \citep{whittaker1988treatise} is used as they claimed.
The fourth Jacobi theta function is defined as $\vartheta_{4}(u, q)=1+2
\sum_{n=1}^{\infty}(-1)^{n} q^{n^{2}} \cos (2\pi n u)$, and the derivative
of $\vartheta_{4}$ with respect to $u$ is $\vartheta^{\prime}_{4}(u,
q)=4\pi \sum_{n=1}^{\infty}n\,(-1)^{n+1} q^{n^{2}} \sin (2\pi n u)$.} As
$I_{1}$ approaches $I_{2}$, the period $T_{1}$ in Eq.~(\ref{eq:phi2})
approaches $2\pi I_{1}/J$. Generally, $T$ and $T_{1}$ are not commensurate
with each other, so the motion of the body is not periodic in the inertia
frame. For simplicity, we define
\citep{Zimmermann:1980ba,VanDenBroeck:2004wj}
%--
\begin{align}
    &\Omega_{\mathrm{p}} \equiv\frac{2 \pi}{T} =\frac{\pi b}{2
    K(m)}\left[\frac{\left(I_{3}-I_{2}\right)\left(I_{3}-I_{1}\right)}{I_{1}
    I_{2}}\right]^{1 / 2}\label{eqn:precession_angular}\,,\\
    &\Omega_{\mathrm{r}}\equiv  \frac{2 \pi}{T_{1}}-\frac{2 \pi}{T} = \frac{J}{I_{1}}+\frac{2\pi
    \mathrm{i}}{T} \frac{\vartheta_{4}^{\prime}(\mathrm{i} \pi \alpha,
    q)}{\vartheta_{4}(\mathrm{i} \pi \alpha, q)}-\Omega_{\mathrm{p}} \label{eqn:rotation_angular}\,,
\end{align}
%--
for later use.

%---------------------------------------------------------------
\subsection{Numerical approach using quaternions}
\label{sec:numerical_solutions}
%---------------------------------------------------------------

Although the analytical solution given in the above subsection is exact,
the use of it is not intuitive. Here we discuss a numerical method to
integrate the equations of motion. There are two reasons for the need of a
numerical method. First, numerical methods can avoid the use of the
elliptic functions. Second, they can be easily applied to general equations
of motion where precessions are much involved with torques in consideration.
In generic cases with torques, analytical solutions usually do not exist.

In our numerical calculation, we employ the numerical method solving
3-dimensional rotations through the use of quaternions, which is a
mathematically equivalent formalism to the aforementioned one using the
Euler angles in describing rotations in the 3-dimensional space
\citep{arribas2006quaternions}. The rotation matrix can be either written
in terms of trigonometric functions of the Euler angles, or expressed by a
specific quaternion whose time evolution is determined by the angular
velocity of the rigid body. In numerical integrations, the latter is
preferred because it produces stable results more efficiently
\citep{arribas2006quaternions}.

A quaternion, $q = q_0 + q_{1}\mathbf{i} + q_{2}\mathbf{j}+q_{3}\mathbf{k}
$, in the quaternion basis, $\{ 1,\, \mathbf{i},\, \mathbf{j},\,\mathbf{k}
\}$, is usually denoted as
%--
\begin{equation}
    q=\left(q_{0}, \,\mathbf{q}\right)\,,
\end{equation}
%--
with $\mathbf{q} = (q_1, \, q_2, \, q_3)$ being a 3-vector when used in
calculating 3-dimensional rotations. The rotation transformation from
$\mathbf{r}$ to $\mathbf{r^{\prime}}$ is performed via
\citep{coutsias2004quaternions}
%--
\begin{align}
\label{eqn:quaternion}
    (0, \mathbf{r^{\prime}}) &=q\left(0,\mathbf{r}\right)\tilde{q}
    \nonumber\\
    &=\left(0,\big(q_{0}^{2}-\mathbf{q} \cdot \mathbf{q}\big)
    \mathbf{r}+2 \mathbf{q}(\mathbf{q} \cdot \mathbf{r})+2 q_{0} \mathbf{q}
    \times \mathbf{r} \right)\nonumber\\
    &=\left(0,\mathcal{R}\,\mathbf{r}\right)\, ,
\end{align}
%--
where $\tilde{q}=\left(q_{0},\mathbf{-q}\right)$ is the conjugate
quaternion of $q$. Note that from the first to the second line we have used
the multiplication rule of quaternions, and from the second to the
third line the usual dot product and cross product of 3-vectors are
applied. Explicitly, the rotation matrix $\mathcal{R}$ in
Eq.~(\ref{eqn:quaternion}) is
%--
\begin{equation}
\label{eqn:rotation}
    {\mathcal{R}=}\left(\begin{array}{ccc}
    q_{0}^{2}+q_{1}^{2}-q_{2}^{2}-q_{3}^{2} & 2q_{1} q_{2}-2q_{0} q_{3} &
    2q_{1} q_{3} + 2q_{0} q_{2} \\
    2q_{1} q_{2} + 2q_{0} q_{3} & q_{0}^{2}-q_{1}^{2}+q_{2}^{2}-q_{3}^{2} &
    2q_{2} q_{3} - 2q_{0} q_{1} \\
    2q_{1} q_{3} - 2q_{0} q_{2} & 2q_{2} q_{3} + 2q_{0} q_{1} &
    q_{0}^{2}-q_{1}^{2}-q_{2}^{2} + q_{3}^{2}
    \end{array}\right)\,,
\end{equation}
%--
which equals the normal Euler rotation matrix. Comparing the elements of
them, we can relate the quaternion $q$ with the Euler angles through
%--
\begin{align}
    &{q_{0}=\cos \frac{\theta}{2}  \cos
    \left(\frac{1}{2}(\phi+\psi)\right)} \,, \label{eqn:euler_to_qua1} \\
    &{q_{1}=\sin \frac{\theta}{2}  \cos
    \left(\frac{1}{2}(\phi-\psi)\right)} \,, \label{eqn:euler_to_qua2}\\
    &{q_{2}=\sin \frac{\theta}{2}  \sin
    \left(\frac{1}{2}(\phi-\psi)\right)} \,, \label{eqn:euler_to_qua3}\\
    &{q_{3}=\cos \frac{\theta}{2}  \sin
    \left(\frac{1}{2}(\phi+\psi)\right)}\,.
\label{eqn:euler_to_qua4}
\end{align}
%--
In addition, the differential equation that governs the time evolution of
the quaternion $q$ can be expressed as \citep{coutsias2004quaternions,
betsch2009rigid}
%--
\begin{equation}
\label{eqn:diff_quaternion}
    \frac{\rmd q}{\rmd t}=\frac{1}{2} q
    \left(0,\boldsymbol{\omega}\right)=\frac{1}{2}
    \left(\begin{array}{cccc}
    {0} & {-\omega_{1}} & {-\omega_{2}} & {-\omega_{3}} \\
    {\omega_{1}} & {0} & {\omega_{3}} & {-\omega_{2}} \\
    {\omega_{2}} & {-\omega_{3}} & {0} & {\omega_{1}} \\
    {\omega_{3}} & {\omega_{2}} & {-\omega_{1}} & {0}
    \end{array}\right)
    \left(\begin{array}{l}
    {q_{0}} \\
    {q_{1}} \\
    {q_{2}} \\
    {q_{3}} 
    \end{array}\right)\,.
\end{equation}
%--

%--
\begin{figure}
    \centering
    \includegraphics[width=8cm]{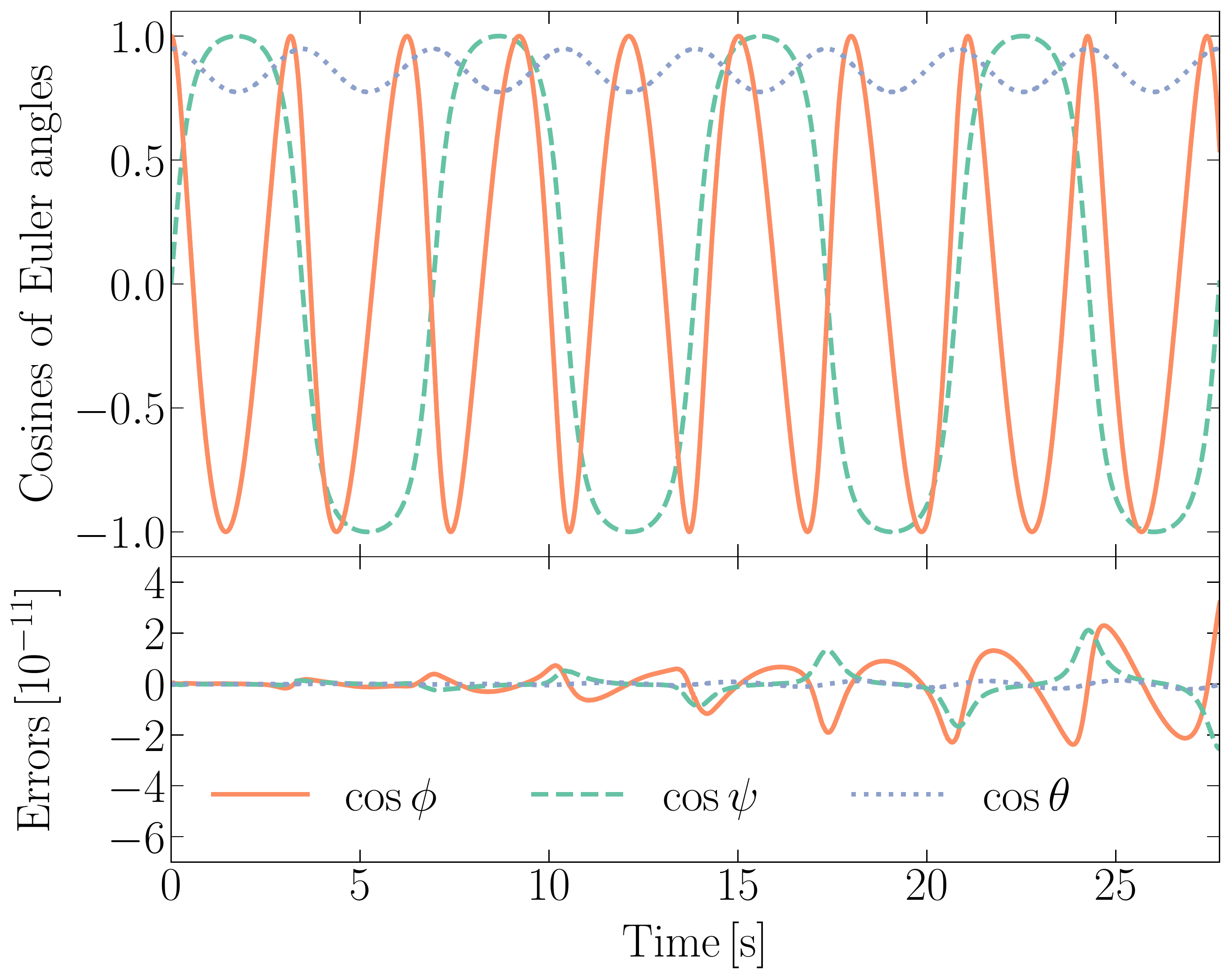}
    \caption{ ({\it Upper}) The numerically solved time evolution for
    cosines of $\theta$, $\psi$, and $\phi$. ({\it Lower}) Absolute error
    of the numerical results with respect to the analytical results for
    cosines displayed in the upper panel. In this plot, for illustrative
    purposes we have taken generic values for the rigid body:
    $I_{1}/I_{3}=1/3$, $I_{2}/I_{3}=2/3$, and
    $a=b=1\,\textrm{rad\,s}^{-1}$. With these values, we have a
    free-precession period $T=6.9\,\textrm{s}$. \Reply{The figure contains four free precession periods.}}
    \label{fig.dynamics_example}
\end{figure}
%--

Once the initial orientation of the rigid body is given, one can translate
it into the initial value of the quaternion $q$ using
Eqs.~(\ref{eqn:euler_to_qua1}--\ref{eqn:euler_to_qua4}). Together with the
initial values of $\omega_1$, $\omega_2$, and $\omega_3$,
Eqs.~(\ref{eqn:euler_body1}--\ref{eqn:euler_body3}) and
Eq.~(\ref{eqn:diff_quaternion}) can be integrated to obtain the angular
velocities in the body frame and the time evolution of the quaternion
$q(t)$, hence the elements of the rotation matrix $\mathcal{R}$ at any
given time. The Euler angles can be recovered by the elements of the matrix
$\mathcal{R}$ via
%--
\begin{align}
    &{\phi=\tan
    ^{-1}\left(-\frac{\mathcal{R}_{13}}{\mathcal{R}_{23}}\right)} \,,\\
    &{\theta=\tan
    ^{-1}\left(\frac{\sqrt{1-\mathcal{R}_{33}^{2}}}{\mathcal{R}_{33}}\right)}
    \,,\\
    &{\psi=\tan
    ^{-1}\left(\frac{\mathcal{R}_{31}}{\mathcal{R}_{32}}\right)}\,,
\end{align}
%--
where $\mathcal{R}_{ij}$ $(i,j = 1,2,3)$ is the component of the matrix
$\mathcal{R}$. In Fig.~\ref{fig.dynamics_example} we present an explicit
example, where the numerical result and the absolute error relative to the
analytical solution are shown. Without dedicated efforts in obtaining
Fig.~\ref{fig.dynamics_example}, the numerical accuracy is already well
below a few parts in trillion in this case. Note that this example is for a
very generic case of free precession for a triaxially deformed rigid body,
while the realistic situation for NSs is much milder as we are to discuss
below, thus we expect significantly better numerical accuracy than what is
shown in Fig.~\ref{fig.dynamics_example}. 

%---------------------------------------------------------------------
\subsection{Physical parameters for triaxial NSs}
\label{sec:dynamics_NS}
%---------------------------------------------------------------------

Before closing this section, let us discuss some typical values for
relevant physical parameters of triaxial NSs. To describe the precession of
triaxial NSs, one can define three small parameters out of $I_1, \, I_2,\,
I_3, \, a$, and $b$, appearing in the previous equations. They are usually
taken as the oblateness
%--
\begin{equation}
    \epsilon \equiv \frac{I_{3}-I_{1}}{I_{3}}\,,
\end{equation}
%--
the nonaxisymmtry
%--
\begin{equation}
    \delta \equiv \frac{I_{2}-I_{1}}{I_{3}-I_{2}} \, ,
\end{equation}
%--
and the tangent of the initial wobble angle
%--
\begin{equation}
    \gamma \equiv \tan \theta_{\rm min} = \frac{I_{1} a}{I_{3} b}\, .
\end{equation}
%--

We first discuss the oblateness due to elastic deformations of NSs. For a
conventional NS with a liquid core and a solid crust,
the oblateness is \citep{1971AnPhy..66..816B,Jones:2000ud},
%--
\begin{equation}
\label{eqn:epsilon}
    \epsilon_{\mathrm{elast}}=\eta \,\epsilon_{0}\,,
\end{equation}
%--
where $\eta$ is the so-called rigidity parameter \citep{Jones:2000ud}, and
$\epsilon_0$ is the zero-strain oblateness. The rigidity parameter $\eta$
is unity for a perfectly rigid star and zero for a liquid star. By assuming
the shear modulus in the crust to be constant, the rigidity parameter for a
NS with a liquid core can be approximated as \citep{1971AnPhy..66..816B}
%--
\begin{equation}
\label{eqn:rigidity}
    \eta \simeq \frac{57 \mu V_{\mathrm{c}}}{10 |E_{g}|} \simeq 2.3 \times 10^{-5}
    \left(\frac{\mu}{ 10^{30} \, \mathrm{erg \, cm^{-3}}}\right)
    R_{6}^{4}\,M_{1.4}^{-2}\,,
\end{equation}
%--
where $\mu$ is the shear modulus of the crust, $V_{\textrm{c}}$ is the volume of the crust, 
and $E_{g}=- 3GM^{2}/5R$ is the gravitational binding energy of the NS. The notations
$M_{1.4}$ and $R_{6}$ represent the dimensionless NS mass $M_{1.4} \equiv
M/\left(1.4\,M_{\odot}\right)$ and the dimensionless NS radius $R_6 \equiv
R/(10^{6}\,\mathrm{cm})$. \citet{Cutler:2002np} adopted a relativistic NS
structure and solved for the strain field in the crust which evolves as the
NS spins down. They found that $\eta$ is smaller than the estimation in
\citet{1971AnPhy..66..816B} by a factor of $\sim 40$. The estimation of the
zero-strain oblateness $\epsilon_{0}$ is
\citep{Cutler:2002np,VanDenBroeck:2004wj},
%--
\begin{equation}
\label{eqn:ep0}
    \epsilon_{0} \simeq \frac{ \Omega_{\rm r}^{2} R^{3} }{G M} =
    2.1\times10^{-3}\left(\frac{f_{\rm r}}{100\,
    \mathrm{Hz}}\right)^{2}R_{6}^{3}M_{1.4}^{-1}\,,
\end{equation}
%--
where $f_{\rm r} = \Omega_{\rm r}/2\pi $ is the spin frequency of
the NS, and $\Omega_{\rm r}$ is defined in Eq.~(\ref{eqn:rotation_angular}).
Combining Eqs.~(\ref{eqn:epsilon}--\ref{eqn:ep0}), we obtain the
oblateness due to the elastic deformation
%--
\begin{equation}
\label{eqn:oblateness}
    \epsilon_{\mathrm{elast}} \simeq 4.9 \times 10^{-8} \left(\frac{f_{\rm
    r}}{100\, \mathrm{Hz}}\right)^{2} \left(\frac{\mu}{ 10^{30} \,
    \mathrm{erg \, cm^{-3}}}\right) R_{6}^{7}\,M_{1.4}^{-3} \,.
\end{equation} 
%--

The oblateness for a NS due to elastic deformation is also limited by the
breaking strain $\sigma_{\textrm{break}}$. According to
\citet{Owen:2005fn}, the largest oblateness is
%--
\begin{align}
    \epsilon_{\mathrm{max} } = 3.4 \times
    10^{-7}\left(\frac{\sigma_{\mathrm{break} }}{10^{-2}}\right) \frac{
    M_{1.4}^{-2.2}\,R_{6}^{4.26} }{1+0.7M_{1.4}R_{6}^{-1}} \,.
\end{align} 
%--
The value of $\sigma_{\textrm{break}}$ is uncertain. Early estimations are
in the range from $10^{-4}$ to $10^{-2}$ \citep{ruderman1992structure}.
\citet{Horowitz:2009ya} found that $\sigma_{\textrm{break}}$ is
around 0.1 by simulating the crust as Coulomb solids. \Reply{Recent
semi-analytical lattice studies of \citet{Baiko:2018jax} showed that
$\sigma_{\textrm{break}}$ is more like to be 0.04. Note that with this
value of $\sigma_{\textrm{break}}$, the largest oblateness is about
$\epsilon_{\textrm{max}}=8\times 10^{-7} $.}

Now we turn to the nonaxisymmetry. NSs are biaxial when $\delta$ is zero or
infinity, and triaxial when $\delta$ has a finite value. Due to the complex
evolution and relaxation of the crust after the star's birth and during the
accretion in the late lifetime \citep{Link:1998km, Shakura:1998ps, Link:2002jk,
Akgun:2005nd}, deformed NSs are possible to be triaxial, characterized by
finite values of $\delta$. The magnetic stresses might contribute to the
triaxility as well \citep{Wasserman:2002ec}. The nonaxisymmetry depends on
the evolution and relaxation of the crust and the magnetic stresses, which
are complex, especially during dynamical or explosive processes. Because of
our lack of knowledge about $\delta$, a measurement of it would be
particularly exciting.

As for the wobble angle $\theta$, there is no physical limitation for
slowly rotating NSs. However, for a fast rotating one, as the rotational
bulge of the NS turns larger, more matter needs to be displaced during the
precession, leading to a larger crust strain
\citep{Jones:2000ud,VanDenBroeck:2004wj}. In order to keep the strain below
the limit of $\sigma_{\textrm{break}}$, the star can only possess a small
wobble angle. \citet{Jones:2000ud} estimated the maximum allowed wobble
angle,
%--
\begin{equation}
\label{eqn:theta_constrain}
    \theta_{\textrm{max}} \approx 0.45\left(\frac{100 \,
    \mathrm{Hz}}{f_{\rm
    r}}\right)^{2}\left(\frac{\sigma_{\mathrm{break}}}{10^{-3}}\right)
    M_{1.4}R_{6}^{-3}\,.
\end{equation}
%--
The constraint on the wobble angle depends on the rotation frequency and
the breaking strain $\sigma_{\textrm{break}}$. For a NS with $f_{\rm r}=100
\, \textrm{Hz}$ and $\sigma_{\textrm{break}}=10^{-3}$, the wobble angle is
smaller than $0.45 \,\textrm{radians}$. If we take the breaking strain in
the extreme case where $\sigma_{\textrm{break}}=0.1$, the wobble angle is
basically unlimited even for a fast rotating NS at a spin frequency of
$f_{\rm r}=500\,\textrm{Hz}$.

For the theoretical analysis in subsequent sections, we apply series
expansion for the trigonometric functions of the three Euler angles in
Section~\ref{sec:analy_solution}, assuming a small oblateness, a small
nonaxisymmetry, and a small wobble angle. The benefit of the perturbative treatment is the
great simplification it brings and the explicit harmonics appearing in the spectra. As for generic cases when one or more of
these parameters are large, one can always restore back to the exact
solution (or the numerical scheme) for a careful check.

Following \citet{Zimmermann:1980ba} and \citet{VanDenBroeck:2004wj}, we
find that practically, it is more convenient to use
%--
\begin{align}
    &\kappa \equiv \frac{1}{16}\frac{I_{3}}{I_{1}}
    \frac{I_{2}-I_{1}}{I_{3}-I_{2}}\, ,
\end{align}
%--
than $\delta$ in the expansion. Up
to the leading order, $\kappa$ and $\delta$ are related by $\kappa \simeq
\delta/16$. A constant of 1/16 is included for the convenience of later
computation \citep{VanDenBroeck:2004wj}. The parameter $m$ in
Eq.~(\ref{eqn:modulus}) is then simplified to $m = 16 \kappa \gamma^{2}$.
In the series expansion, different from \citet{VanDenBroeck:2004wj}, we
treat $\gamma$ and $\kappa$ independently and do not assume any
hierarchy between them. The series expansions of trigonometric functions of
the three Euler angles up to the second order of $\gamma$ and $\kappa$ are,
%--
\begin{align}
    \label{eq:series:begin}
    \cos \phi =& \cos
    \left[(\Omega_{\mathrm{r}}+\Omega_{\mathrm{p}})t\right]\,,\\
    \sin \phi =& \sin
    \left[(\Omega_{\mathrm{r}}+\Omega_{\mathrm{p}})t\right]\,,\\
    \cos \theta =& 1-\frac{\gamma ^2}{2} \,,\\
    \sin \theta =& \gamma +8 \gamma \kappa \sin^2 \left(\Omega_{\mathrm{p}}
    t\right)\,,\\
    \cos \psi =& \sin \left(\Omega_{\mathrm{p}} t \right) \left[1+(8\kappa
    + 32 \kappa^{2})\cos ^{2} \left( \Omega_{\mathrm{p}}t
    \right)\right]\nonumber \\
    & + \sin \left(\Omega_{\mathrm{p}} t \right)\left[16 \kappa ^2 \cos
    ^2\left(\Omega_{\mathrm{p}} t\right) \left(3 \cos \left(2
    \Omega_{\mathrm{p}} t\right)+1\right)\right]\,,\\
    \sin \psi =& \cos \left(\Omega_{\mathrm{p}} t \right)\left[1-(8\kappa +
    32 \kappa^{2})\sin ^{2} \left( \Omega_{\mathrm{p}}t
    \right)\right]\nonumber \\
    & - \cos \left(\Omega_{\mathrm{p}} t \right)\left[96 \kappa ^2 \sin
    ^2\left(\Omega_{\mathrm{p}} t\right) \cos ^2\left( \Omega_{\mathrm{p}}
    t\right)\right]\,.
    \label{eq:series:end}
\end{align}

\Reply{In subsequent sections, we apply the analytical solution in
Eqs.~(\ref{eqn:omega1}--\ref{eq:phi2}) and above estimation for the
oblateness, the nonaxisymmetry, and the wobble angle to investigate the
timing behavior and the GW radiation of precessing triaxial NSs. Note that
the internal, fluid-dynamical mechanisms that might reorient the NS over
time are not included in the calculation. They should exist in general, but
depend on the equation of state of NSs \citep[see e.g.][for solid strange
stars]{Xu:2003xe}. In addition, the magnetic field evolution or pulsation
will lead to time-dependent changes in the oblateness and the
nonaxisymmetry as well.}

\Reply{The braking torques \citep{1970ApJ...160L..11G} and the
gravitational radiation reaction \citep{Cutler:2000bp} are also omitted for
simplicity, which will make the spin frequency decrease and change the
wobble angle. For very young and/or highly-magnetized NSs, this might be
quite important. Magnetic pressures can significantly adjust the precession
period and lead to a nonzero time derivative for the free precession
angular frequency, $\dot{\Omega}_{\rm p}$, because the oblateness
might evolve rapidly and the NS spins down much faster due to large
spin-down torques \citep{Zanazzi:2020vyp, Levin:2020rhj, Suvorov:2020eji}.
For some systems, the braking torques and the gravitational radiation
reaction may become important over long timescales. The details are beyond
the scope of this paper and deserve further investigation.}

%---------------------------------------------------------------------
\section{Modulated timing and pulse signals}
\label{sec:pulsar_radiation}
%---------------------------------------------------------------------

If a triaxially-deformed freely-precessing NS is observed as a pulsar, the
free precession will introduce characteristic modulations on the timing and
pulse signals. These modulations might be revealed by radio and/or X-ray
observations. In Section~\ref{sec:phase}, we discuss the phase modulations
of precessing triaxial NSs and show the residuals of spin period and spin
period derivative for different initial values of the wobble angle. In
Section~\ref{sec:pulse_width}, the pulse-width modulations for different
initial values of the wobble angle are displayed.

%---------------------------------------------------------------------
\subsection{Phase modulation}
\label{sec:phase}
%---------------------------------------------------------------------

\begin{figure}
    \centering
    \includegraphics[width=8cm]{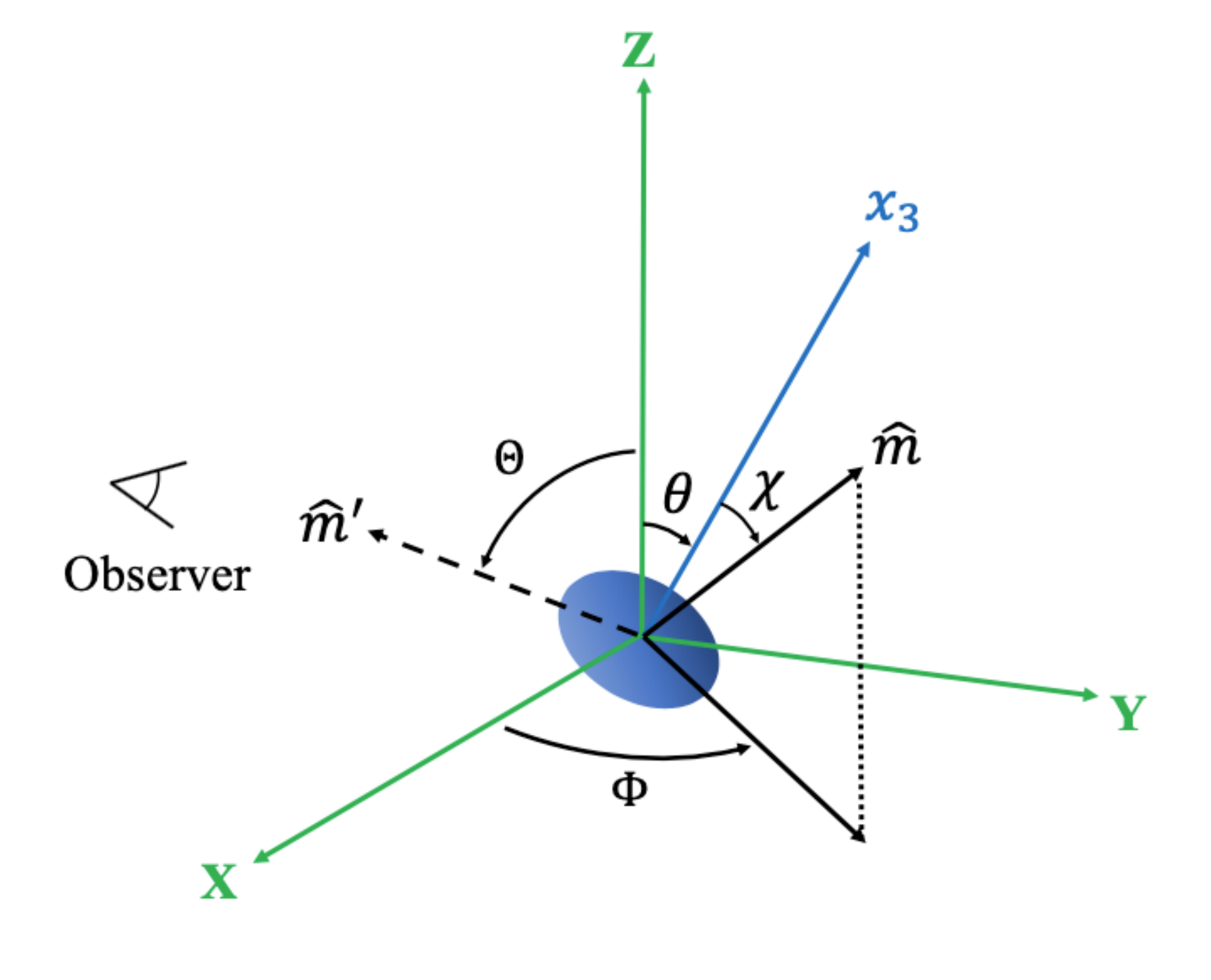}
    \caption{Geometry of a freely precessing triaxial pulsar. The observer
    is in the $\textrm{X}$-$\textrm{Z}$ plane. The dipole moment is denoted
    as $\widehat{m}$, and the dashed line $\widehat{m}^{\prime}$ represents
    the dipole moment when it sweeps through the $\textrm{X}$-$\textrm{Z}$
    plane, namely the moment when the observer can see the pulse. We denote
    the angle between $\widehat{e}_{\rm Z}$ and $\widehat{m}$ as $\Theta$,
    and the angle between $\widehat{e}_{3}$ and $\widehat{m}$ (i.e. the
    magnetic inclination angle) as $\chi$.}
    \label{fig.timing}
\end{figure}

Following \citet{Jones:2000ud}, we assume for simplicity that the pulsar
beam is in the same direction as the magnetic dipole moment $\widehat{m}$.
Once the dipole moment sweeps through the plane defined by the line of
sight and the spin angular momentum, a pulse can be observed. In
Fig.~\ref{fig.timing}, we show the geometry of a freely
precessing triaxial NS. We denote the polar angle between
$\widehat{e}_{\textrm{Z}}$ and $\widehat{m}$ as $\Theta$. We denote the
azimuthal angle between $\widehat{e}_{\textrm{X}}$ and the projection of
$\widehat{m}$ on the $\textrm{X}$-$\textrm{Y}$ plane as $\Phi$. It is
related to the Euler angles via \citep{Jones:2000ud},
%--
\begin{equation}
    \Phi=\phi-\frac{\pi}{2}+\arctan \left(\frac{\cos \psi \sin \chi}{\sin
    \theta \cos \chi-\sin \psi \sin \chi \cos \theta}\right)\,,
\end{equation}
%--
where $\chi$ is the magnetic inclination angle between $\widehat{e}_{3}$
and $\widehat{m}$.

The time derivative of $\Phi$ is the instantaneous spin angular
frequency of the NS. Its time-averaged value corresponds to the mean spin
angular frequency obtained in the observation. Note that for a precessing
triaxial NS, all of the three Euler angles change with time. Especially,
the wobble angle varies in a range from its minimum value $\theta_{\rm{min}}$ to
the maximum, $\theta_{\rm{max}}$. We discuss separately the two situations for
$\theta_{\rm{min}} > \chi$ and $\theta_{\rm{max}} < \chi$ to obtain the
timing residual of the precessing NS.
%--
\begin{itemize}
    \item When $\theta_{\rm{min}} > \chi$, the time-averaged spin frequency
    of the NS is $\dot{\phi}$. Therefore, the precession-induced phase
    residual is \citep{Jones:2000ud}
    %--
    \begin{align}
    \label{eqn:residual1}
        \Delta \Phi =\Phi-\left(\phi-\frac{\pi}{2}\right) =\arctan
        \left(\frac{\cos \psi \sin \chi}{\sin \theta \cos \chi-\sin \psi
        \sin \chi \cos \theta}\right)\,.
    \end{align}
    %--
    \item When $\theta_{\rm{max}} < \chi$, the time-averaged spin frequency
    is $\dot \phi +\dot \psi$, and the precession-induced phase residual is
    \citep{Jones:2000ud}
    %--
    \begin{align}
    \label{eqn:residual2}
        \Delta \Phi &=\Phi-(\phi+\psi) \nonumber\\
        &=\arctan \left[\frac{(\cos \theta-1) \sin \psi \sin \chi-\sin \theta
        \cos \chi}{\cos \psi \sin \chi+(\cos \theta \sin \psi \sin
        \chi-\sin \theta \cos \chi) \tan \psi}\right] \,.
    \end{align}
    %--
\end{itemize}
%--

\begin{figure}
    \centering
    \includegraphics[width=8.2cm]{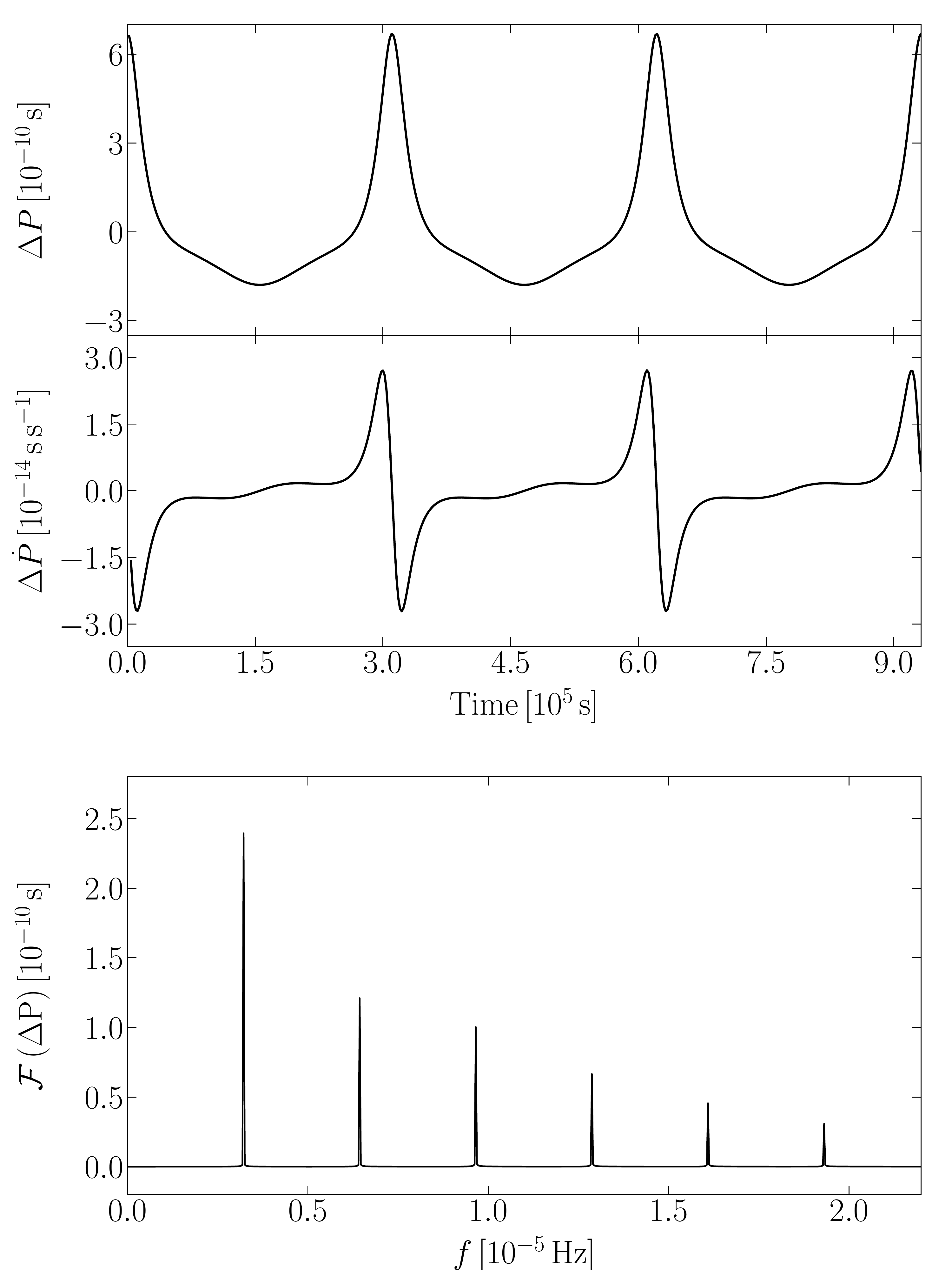}
    \caption{ ({\it Upper}) Precession-induced residuals of the spin period
    and the spin period derivative for a large wobble angle where
    $\theta_{\rm{min}} > \chi$. ({\it Lower}) The Fourier amplitude for the
    spin period residual. In this figure, we have chosen the magnetic
    inclination angle $\chi=\pi/6$, the oblateness $\epsilon=4.9 \times
    10^{-8}$, the nonaxisymmetry parameter $\delta=0.1$, and a wobble angle
    in the range $\theta \in \left(0.79,\,0.84\right)$. With these
    parameters, we have $T_{1}=0.010\,\textrm{s}$, and a free precession
    period $T=3.1\times10^{5}\, \textrm{s}$.
    }\label{fig.timing_large}
\end{figure}

\begin{figure}
    \centering
    \includegraphics[width=8.2cm]{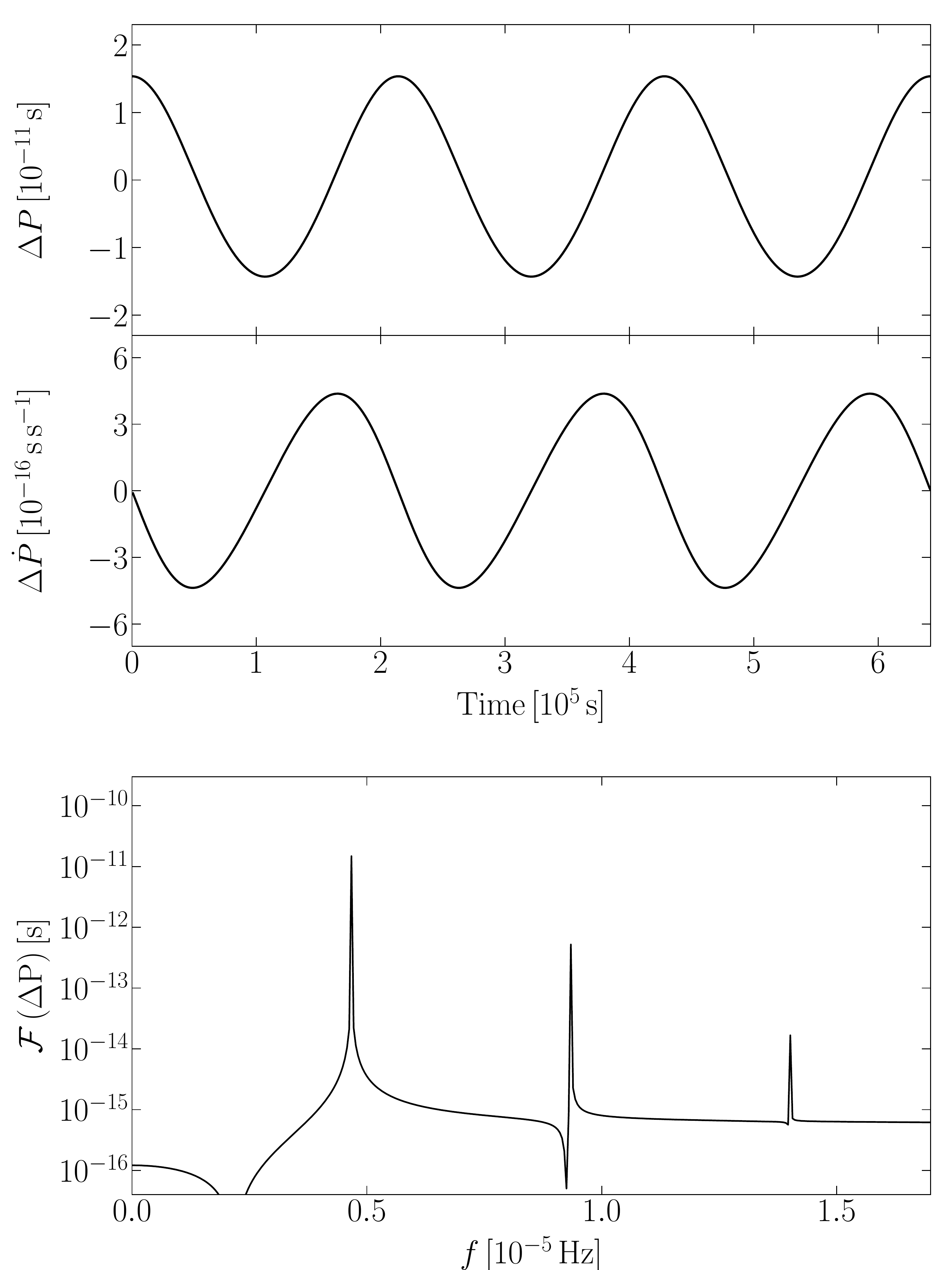}
    \caption{ ({\it Upper}) Precession-induced residuals of the spin period
    and the spin period derivative for a small wobble angle where
    $\theta_{\rm{max}} \ll \chi$. ({\it Lower}) The Fourier amplitude for
    the spin period residual. In this figure, we have chosen the same
    $\chi$, $\epsilon$, and $\delta$ as in Fig.~\ref{fig.timing_large}, but
    a small wobble angle in the range $\theta \in
    \left(0.017,\,0.018\right)$. With these parameters, we have
    $T_{1}=0.010\,\textrm{s}$, and a free precession period
    $T=2.1\times10^{5}\,\textrm{s}$. }\label{fig.timing_small}
\end{figure}

The precession-induced residuals of the spin period, $P$, and the spin
period derivative, $\dot P$, can be calculated using the time derivatives
of the precession-induced phase residual,
%--
\begin{align}
\label{eqn:ppdot_modulation1}
    \Delta P &=-\frac{P_{0}^{2}}{2 \pi} \Delta \dot{\Phi}\,, \\
\label{eqn:ppdot_modulation2}
    \Delta \dot{P} &=-\frac{P_{0}^{2}}{2 \pi} \Delta \ddot{\Phi}\,,
\end{align}
%--
where $P_{0}$ is the mean spin period of the NS. Substituting the
time-evolving Euler angles into
Eqs.~(\ref{eqn:residual1}--\ref{eqn:ppdot_modulation2}), one can obtain
the modulations of the spin period and the spin period derivative for
different choices of parameters.

In Fig.~\ref{fig.timing_large} and Fig.~\ref{fig.timing_small} we
respectively present examples for cases of a large wobble angle where
$\theta_{\rm{min}} > \chi$, and a small wobble angle where
$\theta_{\rm{max}} \ll \chi$. In the calculation, we take the magnetic
inclination angle $\chi=\pi/6$, the mean spin period $P_{0}=0.01\,\rm{s}$, 
and we make use of Eq.~(\ref{eqn:oblateness}) to estimate the oblateness. 
%--
\begin{itemize}
    \item The residuals of the spin period and the spin period derivative
    for the case of a large wobble angle are displayed in the upper panel
    of Fig.~\ref{fig.timing_large}. The Fourier transformation of the spin
    period residual is shown in the lower panel of
    Fig.~\ref{fig.timing_large}. The spectrum shows strong peaks at
    frequencies $n\,\Omega_{\textrm{p}} / 2\pi$, where $n$ is a positive integer
    number and $\Omega_{\textrm{p}}$ ($\simeq 2.0\times10^{-5}\,{\rm s}^{-1}$) is
    the free precession angular frequency defined in 
    Eq.~(\ref{eqn:precession_angular}).
    %--
    \item For the small wobble angle limit where $\theta_{\rm{max}} \ll
    \chi$, we display the residuals of the spin period and the spin period
    derivative in the upper panel of Fig.~\ref{fig.timing_small}. We also
    take the Fourier transformation of the spin period residual, whose
    amplitude is shown in the lower panel of Fig.~\ref{fig.timing_small}.
    Notice that a logarithmic scale is used for the Fourier amplitude.
    Compared to the case of a large wobble angle, the harmonics at
    $n\Omega_{\textrm{p}}/2\pi$ ($n\geq 2$) are much weaker than the line
    at $\Omega_{\textrm{p}}/2\pi$ (now, $\Omega_{\textrm{p}} \simeq 3.0
    \times 10^{-5} \, {\rm s}^{-1}$) in the case of a small wobble angle.
    In the small wobble angle limit, the precession-induced spin phase
    residual can be approximated as \citep{Jones:2000ud,Link:2001zr}
%--
\begin{equation}
    \Delta \Phi = -\sin \theta \cot \chi \cos \psi - \frac{1}{4}
    \sin^{2}\theta (1+2\cot^{2}\chi)\sin 2\psi \,.
\end{equation}
%--
Applying the series expansion of the Euler angles in
Eqs.~(\ref{eq:series:begin}--\ref{eq:series:end}) and using
Eq.~(\ref{eqn:residual2}), the spin period residual is given by
%--
\begin{align}
\label{eqn:delp}
    \Delta P \approx &\frac{P_{0}^{2}}{2\pi}\Omega_{\mathrm{p}}\gamma (8 \kappa +1)
    \cot \chi \cos \left(\Omega_{\mathrm{p}} t \right) \nonumber \\
    &+\frac{P_{0}^{2}}{4\pi}\Omega_{\mathrm{p}} \gamma^{2}
    \left(1+2\cot^{2}\chi\right)\cos \left(2\Omega_{\mathrm{p}} t \right) \,.
\end{align}
%--
It shows that at the second order of the wobble angle, the modulation of
the spin period includes the first and the second harmonics of the free
precession angular frequency $\Omega_{\textrm{p}}$, corresponding to the
first two peaks in the lower panel of Fig.~\ref{fig.timing_small}. The
third peak comes from higher-order terms that are not included in the
approximation. The residual of the spin period derivative $\Delta \dot{P}$
can be obtained by taking the time derivative of $\Delta{P}$, which gives
%--
\begin{align}
\label{eqn:delpdot}
    \Delta \dot{P} \approx
    &-\frac{P_{0}^{2}}{2\pi}\Omega_{\mathrm{p}}^{2}\gamma (8 \kappa +1)
    \cot \chi \sin \left(\Omega_{\mathrm{p}} t \right) \nonumber \\
    &-\frac{P_{0}^{2}}{2\pi}\Omega_{\mathrm{p}}^{2} \gamma^{2}
    \left(1+2\cot^{2}\chi\right)\sin \left(2\Omega_{\mathrm{p}} t \right) \,.
\end{align} 
%--
When $\kappa=0$, Eqs.~(\ref{eqn:delp}--\ref{eqn:delpdot}) reduce to the 
corresponding results for a precessing biaxial NS \citep{Link:2001zr}.
\Reply{Note that we have ignored the intrinsic spindown of NSs for simplicity in deriving Eqs.~(\ref{eqn:delp}--\ref{eqn:delpdot}).}

\end{itemize}
%--

%---------------------------------------------------------------------
\subsection{Pulse-width modulation}
\label{sec:pulse_width}
%---------------------------------------------------------------------

\begin{figure}
    \centering
    \includegraphics[width=7cm]{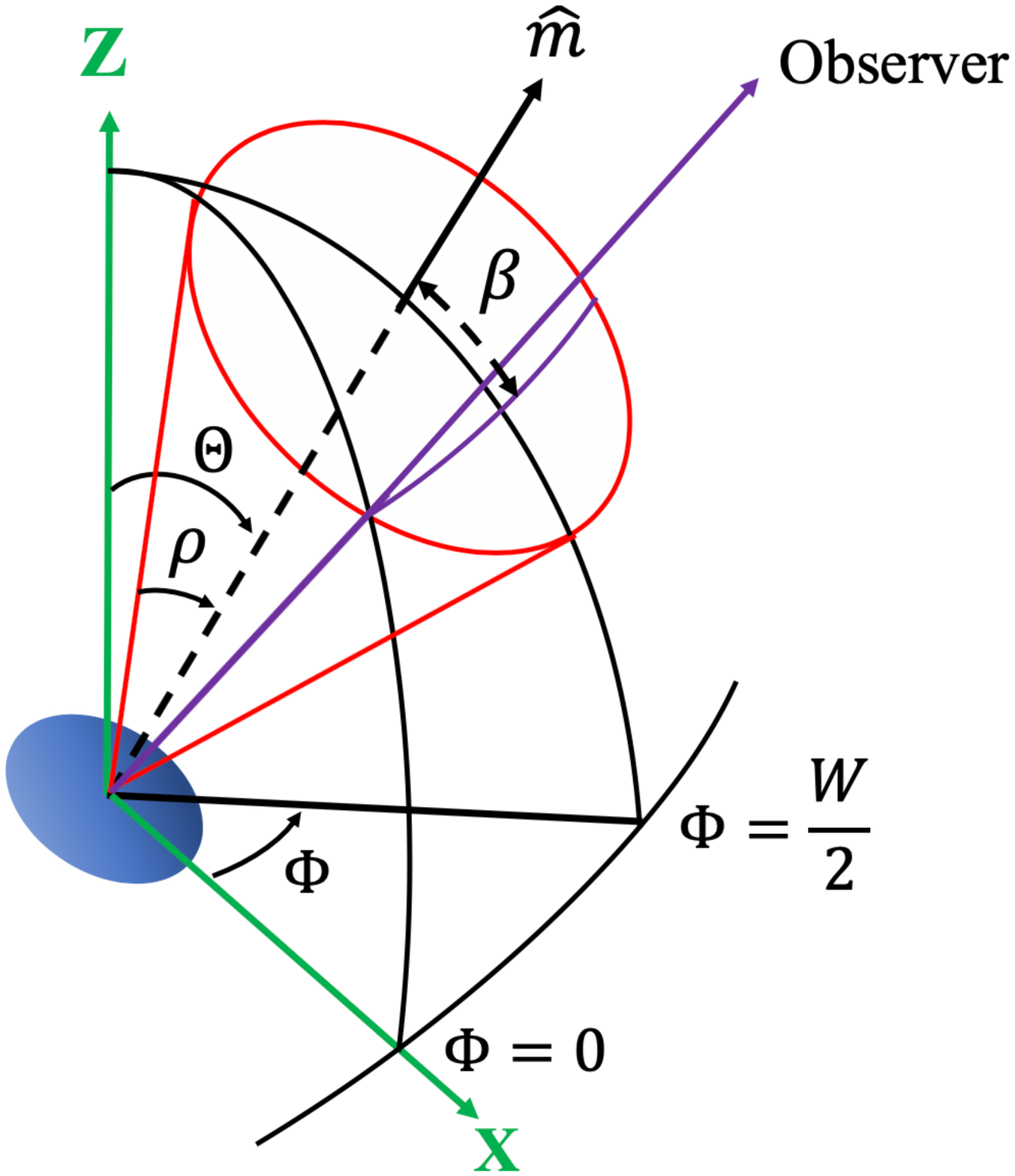}
    \caption{Geometry of the pulsar emission beam in the cone model
    \citep{Gil:1984ads,Lorimer:2005misc}. The emission is confined in a
    cone with an opening angle $\rho$. We denote the impact angle as
    $\beta$, which corresponds to the closest approach between the line 
    of sight and the magnetic dipole moment. Pulse signals can be observed 
    once the line of sight sweeps through the cone. The purple line denotes 
    the sweep of the line of sight, and different cuts of the line of sight 
    through the cone result in different pulse width $W$.}
    \label{fig.pulse_width}
\end{figure}

\begin{figure}
    \centering
    \includegraphics[width=8.2cm]{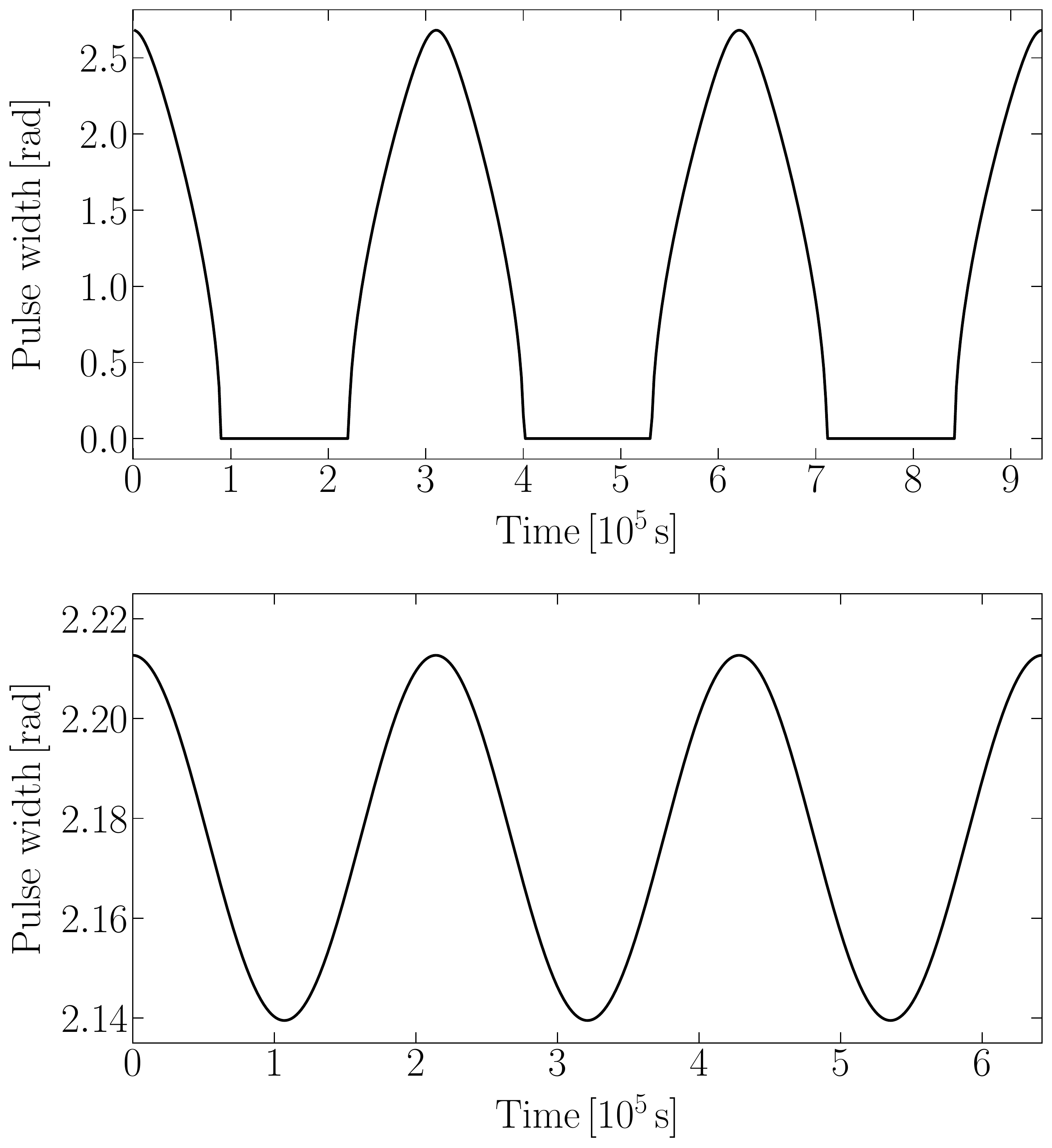}
    \caption{({\it Upper}) The pulse-width modulation in the case of a
    large wobble angle; parameters are the same as in Fig.~\ref{fig.timing_large}. 
    ({\it Lower}) The pulse-width modulation for a small wobble angle; 
    parameters are the same as in Fig.~\ref{fig.timing_small}. 
    For both cases, we have chosen the inclination angle $\iota=5\pi/6$,
    and the angular radius of the emission cone $\rho=\pi/6$.}
    \label{fig.timing_width}
\end{figure}

%--
\begin{figure*}
    \centering
    \includegraphics[width=17cm]{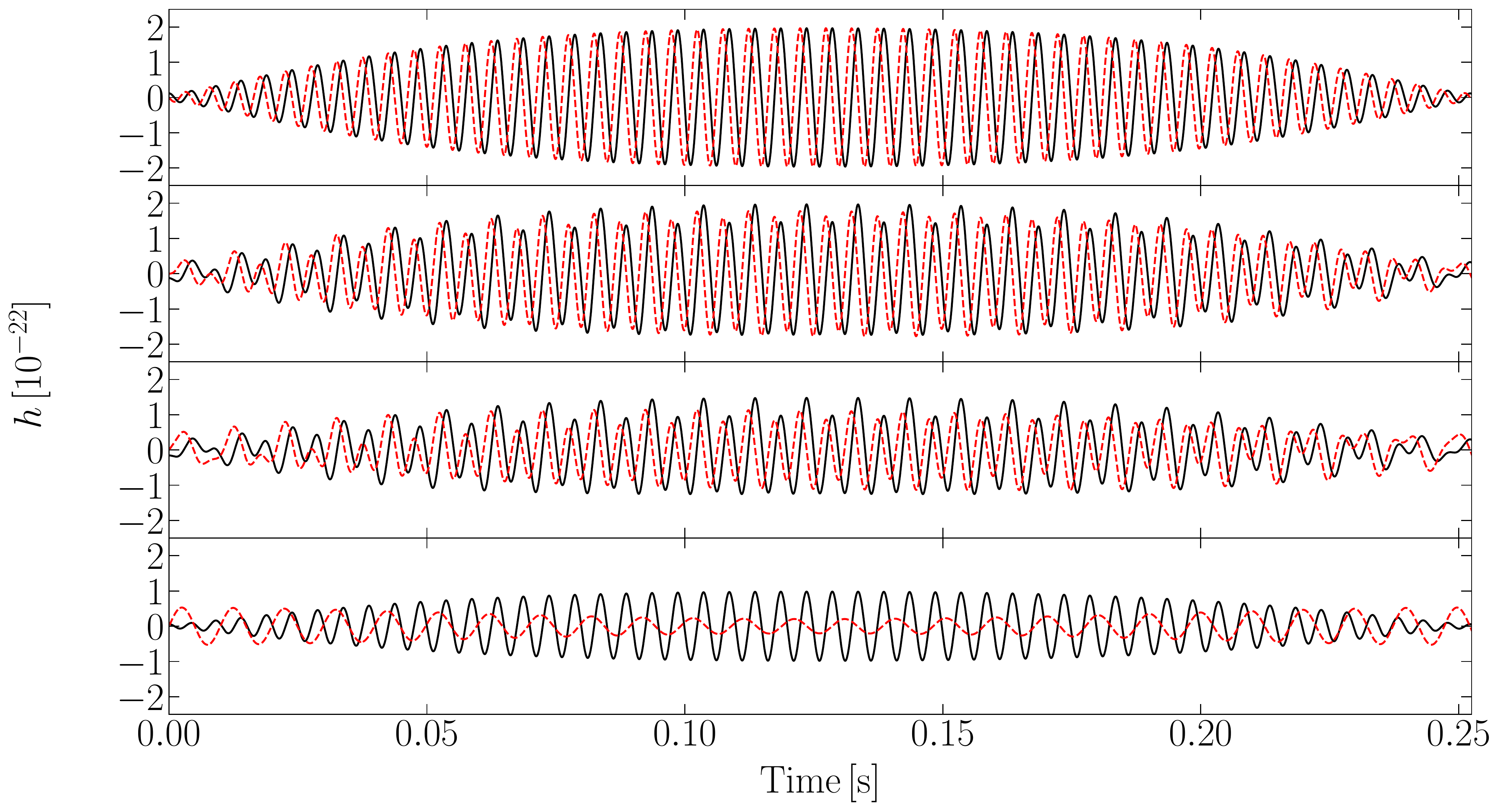}
    \caption{Gravitational waveforms of $h_{+}$ (black solid line) and
    $h_{\times}$ (red dotted line) as a function of time at inclination
    angles (from top to bottom) $\iota=0$, $\pi/6$, $\pi/3$, and $\pi/2$.
    We have chosen the oblateness $\epsilon=0.05$, the nonaxisymmetry
    $\delta=0.8$, the wobble angle in the range $\theta \in
    \left(0.76,\,1.21\right)$, the moment of inertia
    $I_{3}=10^{45}\textrm{g}\,\textrm{cm}^{2}$, and the distance from the
    observer to the source $r=10\,\textrm{kpc}$. With these exaggerated
    parameters for illustrative purposes, we have
    $T_{1}=0.0099\,\textrm{s}$, and a free precession period
    $T=0.50\,\textrm{s}$. This figure contains half of a free precession
    period. }
    \label{fig.gw_large}
\end{figure*}
%--

In order to analyse the pulse-width modulation, we adopt a simple cone
model to describe the radiation of a pulsar. For a more complicated
radiation geometry, our method can be extended as well. In the cone model,
from the geometry in Fig.~\ref{fig.pulse_width} we have \citep{Gil:1984ads,
Lorimer:2005misc},
%--
\begin{align}\label{eq:cone:model}
    \sin ^{2}\left(\frac{W}{4}\right)=\frac{\sin ^{2}(\rho / 2)
    -\sin ^{2}(\beta / 2)}{\sin (\Theta+\beta) \sin \Theta}\,,
\end{align}
%--
where $\Theta$ is defined in Fig.~\ref{fig.timing}, and ${W}$, $\rho$, 
$\beta$ are defined in Fig.~\ref{fig.pulse_width}.

Equation~(\ref{eq:cone:model}) is not exact for the pulse width because in
our case the angle $\Theta$ changes with time. However, as the spin
frequency is much higher than the free precession frequency, the change in
$\Theta$ during a spin period is negligible, thus this approximation is
good enough for our calculation. In this cone model, the observer can
observe the pulse signal once the line of sight enters into the emission
cone. The variation of pulse width can be determined by the angle $\Theta$
once the inclination angle (denoted as $\iota$) between the angular
momentum and the line of sight to the NS is determined. The angle $\Theta$
can be expressed through the Euler angles and the magnetic inclination
angle via \citep{Jones:2000ud,Link:2001zr},
%--
\begin{equation}
    \cos \Theta=\sin \theta \sin \psi \sin \chi+\cos \theta
    \cos \chi \,.
\end{equation}
%--

In Fig.~\ref{fig.timing_width} we present examples for the pulse-width
modulation with a large wobble angle where $\theta_{\rm{min}} > \chi$ and a
small wobble angle where $\theta_{\rm{max}} \ll \chi$. The choices of the
oblateness, the nonaxisymmetry, and the magnetic inclination angle are the
same as in Fig.~\ref{fig.timing_large} and Fig.~\ref{fig.timing_small} for
the large and small wobble angles, respectively. The example for the large
wobble angle is displayed in the upper panel of Fig.~\ref{fig.timing_width}.
For our (extreme) choice of the parameters, the angle between the angular
momentum and the dipole moment changes significantly, $\Theta \in
\left(0.26,\,1.31\right)$. As a consequence, the pulse width changes in a
wide range with $W\in\left(0,2.68\right)$. The line of sight leaves out of the
emission cone due to the free precession during certain time ranges, and then
the pulse width diminishes to zero accordingly. The modulation of pulse
width in the case of a small wobble angle is shown in the lower panel of
Fig.~\ref{fig.timing_width}. In this case, the angle $\Theta$ is in the
range of $\Theta \in \left(0.51,\, 0.54\right)$, and the change of pulse
width is much milder with $W\in\left(2.14, 2.21\right)$.

In the case of a small wobble angle, applying the series expansions of
$\theta$ and $\psi$ in Eqs.~(\ref{eq:series:begin}--\ref{eq:series:end}),
the angle $\Theta$ can be approximated as
%--
\begin{equation}
    \Theta \approx \chi - \sin \theta \sin \psi  \approx \chi - \gamma
    \cos \left(\Omega_{\mathrm{p}} t\right) \,,
\end{equation}
%--
which reduces to the corresponding result for a precessing biaxial NS
\citep{Link:2001zr}.

%---------------------------------------------------------------------
\section{Continuous Gravitational Waves}
\label{sec:gw}
%---------------------------------------------------------------------

We discuss generic continuous GWs from a triaxially-deformed
freely-precessing NS in Section~\ref{sec:g:wf}, and the approximation to
the waveform with small oblateness, a small wobble angle, and small
nonaxisymmetry in Section~\ref{sec:gw_small}. The results mainly follow
\citet{Zimmermann:1980ba} and \citet{VanDenBroeck:2004wj}, but we make
further extensions by assuming no hierarchy in the three small parameters.

%---------------------------------------------------------------------
\subsection{Generic waveform}
\label{sec:g:wf}
%---------------------------------------------------------------------

We use the quadrupole approximation for the continuous GWs from freely
precessing triaxial NSs. In the transverse-traceless (TT) gauge, the metric
perturbation is \citep{Misner:1974qy}
%--
\begin{equation}
\label{eqn:quadrupole}
    h_{i j}^{\mathrm{TT}}=\frac{2G}{r c^{4}} \frac{\rmd^{2} I_{i j}}{\rmd
    \,t^{2}}\,,
\end{equation}
%--
where \Reply{$G$ is the gravitational constant, $c$ is the speed of light, }$I_{ij}$ is the trace-free part of the moment of inertia tensor, and
$r$ is the luminosity distance from the source to the observer.

Alternatively, the quadrupole formula in Eq.~(\ref{eqn:quadrupole}) can be
expressed as \citep{Zimmermann:1980ba}
%--
\begin{equation}
\label{eqn:waveform1}
    h_{i j}^{\mathrm{TT}}=-\frac{2G}{rc^{4}} \mathcal{R}_{i k} \mathcal{R}_{j l
    }A_{kl} \,,
\end{equation}
%--
where $\mathcal{R}$ is the rotation matrix (\ref{eqn:rotation}), and
$A_{kl}$ is determined by the body-frame angular velocities and the
body-frame moments of inertia. For example, we have
%--
\begin{align}
    &A_{11}=2\left(\Delta_{2} \omega_{2}^{2}-\Delta_{3}
    \omega_{3}^{2}\right)\label{eqn:a11}\,,\\
    &A_{12}= \left(\Delta_{1}-\Delta_{2}+\frac{\Delta_{3}^{2}}{I_{3}}\right)\omega_{1}
    \omega_{2} \label{eqn:a12} \,,
\end{align}
%--
where 
%--
\begin{equation}
    \Delta_{1} \equiv I_{2}-I_{3}, \quad \Delta_{2} \equiv I_{3}-I_{1},
    \quad \Delta_{3} \equiv I_{1}-I_{2}\,.
\end{equation} 
%--
The other components of $A_{kl}$ can be obtained from
Eqs.~(\ref{eqn:a11}--\ref{eqn:a12}) by cyclic permutation of the indices.

The waveform in Eq.~(\ref{eqn:waveform1}) is usually decomposed into $h_+$
and $h_\times$,
%--
\begin{equation}
\label{eqn:decompose}
    h_{i j}^{\mathrm{TT}}=  h_{+} \left(\widehat{e}_{+}\right)_{ij} + h_{\times}
    \left(\widehat{e}_{\times}\right)_{ij} \,,
\end{equation}
%--
where $h_{+}$ and $h_{\times}$ represent the radiation of the two
independent polarizations. The polarization tensors, $\widehat{e}_{+}$
and $\widehat{e}_{\times}$, are
%--
\begin{align}
    &\widehat{e}_{+}=\widehat{p} \otimes \widehat{p}-\widehat{q} \otimes
    \widehat{q}\label{eqn:plus}\,,\\
    &\widehat{e}_{\times}=\widehat{p} \otimes \widehat{q}+\widehat{q}
    \otimes \widehat{p} \label{eqn:cross} \,,
\end{align}
%--
where $\widehat{p}$ and $\widehat{q}$ are two unit vectors with
$\widehat{p}\times \widehat{q}$ in the propagation direction of the GWs.
We assume that the observer lies in the \textrm{Y-Z} plane and define the 
inclination angle $\iota$ as the angle between the direction
of the angular momentum $\widehat{e}_{\rm Z}$ and the line of sight to the
NS. In the inertial frame, the unit vectors $\widehat{p}$ and $\widehat{q}$ are
%--
\begin{align}
    &\widehat{p}=-\widehat{e}_{\rm Y} \cos \iota-\widehat{e}_{\rm Z} \sin
    \iota \label{eqn:p}\,,\\
    &\widehat{q}=-\widehat{e}_{\rm X}\label{eqn:q}\,.
\end{align}
%--

Combining Eq.~(\ref{eqn:waveform1}) and
Eqs.~(\ref{eqn:decompose}--\ref{eqn:q}), the waveforms of the two
polarizations are \citep{Zimmermann:1980ba,VanDenBroeck:2004wj}
%--
\begin{align}
    &h_{+} =-\frac{G}{rc^{4}} \big[\left( \mathcal{R}_{2 k} \cos \iota +
    \mathcal{R}_{3 k} \sin \iota\right)\left( \mathcal{R}_{2 l}
    \cos \iota + \mathcal{R}_{3 l} \sin \iota\right) \nonumber \\
    &\quad \quad \quad \quad -\mathcal{R}_{1 k} \mathcal{R}_{1
    l}\big] A_{k l} \label{eqn:waveform_plus}\,,\\
    &h_{\times} =-\frac{2G}{rc^{4}}\left( \mathcal{R}_{2 k} \cos \iota +
    \mathcal{R}_{3 k} \sin \iota\right) \mathcal{R}_{1 l}
    A_{k l}\label{eqn:waveform_cross}\,.
\end{align}
%--

\begin{figure}
    \centering
    \includegraphics[width=8.4cm]{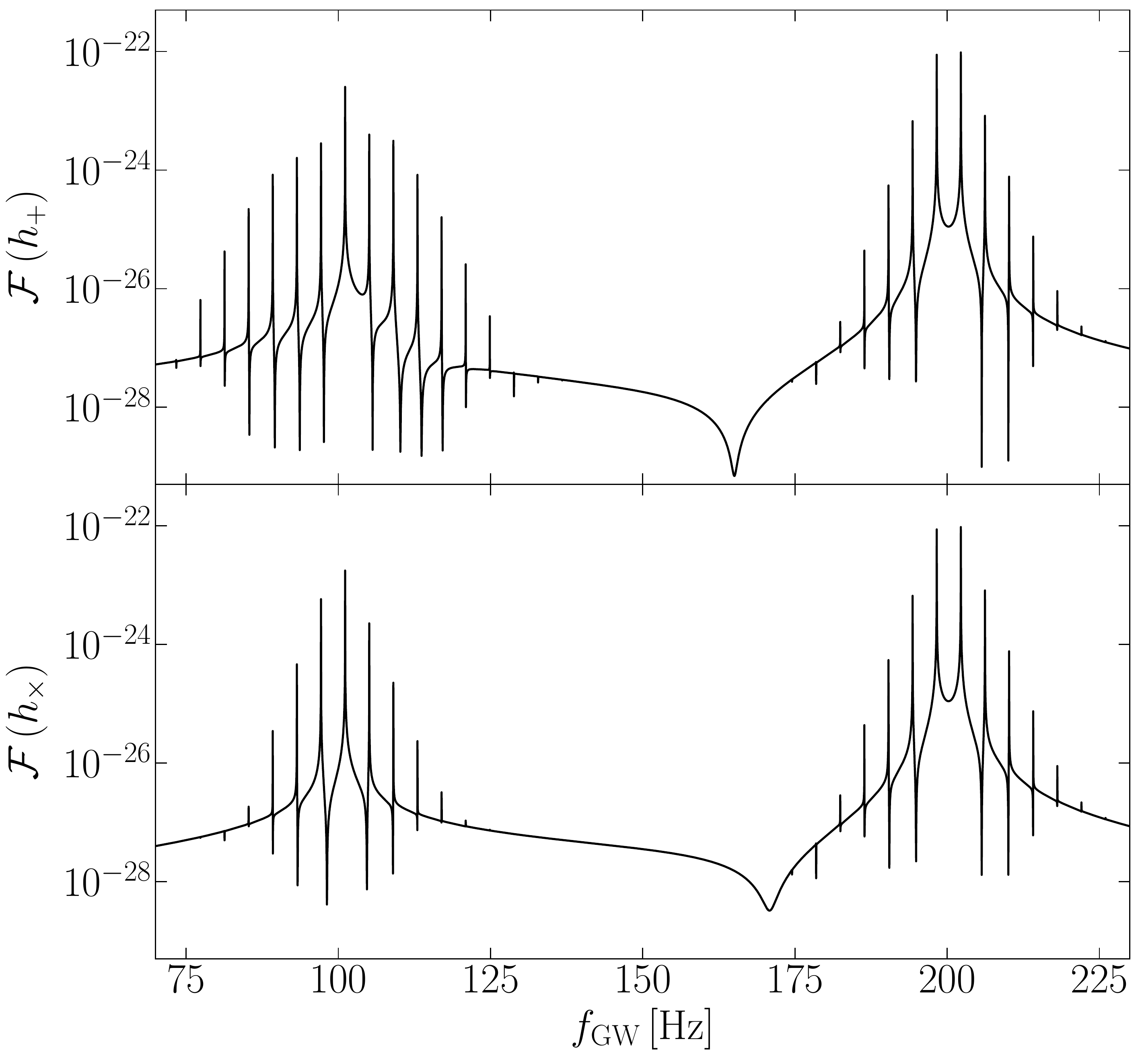}
    \caption{The Fourier amplitudes of $h_{+}$ and $h_{\times}$ for the
    waveform in the second panel of Fig.~\ref{fig.gw_large} with
    $\iota=\pi/6$. } \label{fig.gw_large_fft}
\end{figure}

In Section \ref{sec:free_prec}, we have discussed the time evolution of the
angular velocities in the body frame and the three Euler angles, so we can
obtain $h_{+}$ and $h_{\times}$ at any given time $t$. In
Fig.~\ref{fig.gw_large}, we plot waveforms of $h_{+}$ and $h_{\times}$ in
the time domain at different inclination angles. Note that the parameters
that we have chosen for the plot are exaggerated for NSs for illustrative
purposes. Physical parameters consistent with the estimates in Section
\ref{sec:dynamics_NS} can be easily implemented, but the effects will be
too small for visual inspection. We also show the Fourier
transformation of the waveforms at the inclination angle of $\iota=\pi/6$
in Fig.~\ref{fig.gw_large_fft}. The peaks of the spectra are dominantly at
angular frequencies
%--
\begin{equation}
    \Omega_{\mathrm{r}}+n\,\Omega_{\mathrm{p}}\,,\quad 2\Omega_{\mathrm{r}}+n\,\Omega_{\mathrm{p}}\,,
\end{equation}
%--
where $n=0, \pm1, \pm2, \cdots$ is an integer number.
%--

%---------------------------------------------------------------------
\subsection{Waveform for small oblateness, small wobble angle and small
nonaxisymmetry}
\label{sec:gw_small}
%---------------------------------------------------------------------

Following \citet{Zimmermann:1980ba} and \citet{VanDenBroeck:2004wj}, we
investigate the waveforms in the limit of small oblateness $\epsilon$,
small wobble angle $\theta$, and small nonaxisymmetry $\delta$. The
difference between our work and the previous work is that, instead of
assuming a hierarchy between $\kappa$ and $\gamma$, namely $\kappa \sim
{\cal O}\big( \gamma^2 \big)$ in \citet{VanDenBroeck:2004wj}, we treat
$\gamma$ and $\kappa$ as small quantities independent to each other. As
discussed in Section~\ref{sec:dynamics_NS}, it is more plausible to assume
no intrinsic hierarchy between $\kappa$ and $\gamma$, in particular when
the internal structure of NSs is still rather uncertain.

The procedure to derive the expansion of the waveform is as follows. First
we expand $R_{i j}$ and the angular velocities $\omega_{1}$, $\omega_{2}$,
and $\omega_{3}$ to the second order of $\gamma$ and $\kappa$ using the
expansion of Euler angles in
Eqs.~(\ref{eq:series:begin}--\ref{eq:series:end}). Second, we substitute
the expansion of $R_{i j}$ and $A_{k l}$ into
Eqs.~(\ref{eqn:waveform_plus}--\ref{eqn:waveform_cross}). Third, we retain
the GW waveform to the second order of $\gamma$ and $\kappa$ and combine
the trigonometric functions using trigonometric identities. Such an
extension with independent $\kappa$ and $\gamma$ parameters gives us more
features than what was discovered before.

With the above procedure, we obtain six components of $h_{+}$, which are 
distinguished by different GW frequencies,
%--
\begin{align}
    h_{+}^{1} = & \frac{ - G\epsilon \gamma I_{3} b^{2}\sin 2\iota }{rc^{4}}\cos
    \left[(\Omega_{\mathrm{r}}+\Omega_{\mathrm{p}})t\right]\,,\\
    h_{+}^{2} = & \frac{-32 G\epsilon \kappa I_{3} b^{2} (1+ \cos^{2}
    \iota)}{r c^{4}}\cos \left[2\Omega_{\mathrm{r}}t\right]\,,\\
    h_{+}^{3} = & \frac{2 G\epsilon (64 \kappa^{2} +\gamma^{2})I_{3}
    b^{2}(1+ \cos^{2} \iota)}{r c^{4}}\cos \left[2(\Omega_{\mathrm{r}}+\Omega_{\mathrm{p}})t\right]\,,\\
    h_{+}^{4} = & \frac{- 14G\epsilon \gamma \kappa I_{3}b^{2}\sin 2\iota
    }{r c^{4}}\cos \left[(\Omega_{\mathrm{r}}-\Omega_{\mathrm{p}})t\right]\,,\\
    h_{+}^{5} = & \frac{2G\epsilon \gamma \kappa I_{3} b^{2}\sin 2\iota}{rc^{4}
    }\cos \left[(\Omega_{\mathrm{r}}+3\Omega_{\mathrm{p}})t\right]\,,\\
    h_{+}^{6} = & \frac{-128 G\epsilon \kappa ^{2} I_{3} b^{2}(1+ \cos^{2}
    \iota)}{r c^{4}}\cos \left[2(\Omega_{\mathrm{r}}-\Omega_{\mathrm{p}})t\right]\, .
\end{align}
%--
Similarly, we obtain six components of $h_{\times}$,
%--
\begin{align}
    h_{\times}^{1} = & \frac{ 2G\epsilon \gamma I_{3} b^{2}\sin \iota
    }{r c^{4}}\sin \left[(\Omega_{\mathrm{r}}+\Omega_{\mathrm{p}})t\right]\,,\\
    h_{\times}^{2} = & \frac{64G \epsilon \kappa I_{3} b^{2} \cos \iota}{r c^{4}
    }\sin \left[2\Omega_{\mathrm{r}}t\right]\,,\\
    h_{\times}^{3} = &\frac{-4 G\epsilon (64 \kappa^{2} +\gamma^{2})I_{3}
    b^{2} \cos\iota}{r c^{4}}\sin \left[2(\Omega_{\mathrm{r}}+\Omega_{\mathrm{p}})t\right]\\
    h_{\times}^{4} = & \frac{28G\epsilon \gamma \kappa I_{3}b^{2}\sin \iota
    }{r c^{4}}\sin \left[(\Omega_{\mathrm{r}}-\Omega_{\mathrm{p}})t\right]\,,\\
    h_{\times}^{5} = & \frac{-4G\epsilon \gamma \kappa I_{3} b^{2}\sin
    \iota}{r c^{4}}\sin \left[(\Omega_{\mathrm{r}}+3\Omega_{\mathrm{p}})t\right]\,,\\
    h_{\times}^{6} = & \frac{256 G\epsilon \kappa ^{2} I_{3} b^{2}\cos
    \iota}{r c^{4}}\sin \left[2(\Omega_{\mathrm{r}}-\Omega_{\mathrm{p}})t\right]\,.
\end{align}

Note that the components with frequencies
$\Omega_{\mathrm{r}}+\Omega_{\mathrm{p}}$ and $2\Omega_{\mathrm{r}}$ are
the leading order contributions found in \citet{Zimmermann:1980ba}, and
that the one with frequency $2(\Omega_{\mathrm{r}}+\Omega_{\mathrm{p}})$ is
the third spectral line in \citet{VanDenBroeck:2004wj} once $\kappa$ is
treated as $O(\gamma^2)$.

Below we briefly discuss the waveform for different choices of $\gamma$ and
$\kappa$, and for presentation reasons, we leave their observational
aspects to the next section, together with the possible radio/X-ray
counterparts.
%--
\begin{itemize}
    \item When $\gamma=0$ and $\kappa \neq 0$, the NS does not precess and
    GWs are radiated at twice of the rotation frequency. The radiation is 
    caused by the asymmetry between $I_{1}$ and $I_{2}$.
    \item When $\kappa=0$ and $\gamma \neq 0$, GWs are radiated at 
    $\Omega_{\mathrm{r}}+\Omega_{\mathrm{p}}$ and $2(\Omega_{\mathrm{r}}+\Omega_{\mathrm{p}})$. 
    This is the classical result of a precessing biaxial NS \citep{Zimmermann:1979ip}.
    \item When $\kappa \neq 0$ and $\gamma \neq 0$, the NS is a precessing
    triaxial body. At the first order of $\gamma$ and $\kappa$, continuous
    GWs are emitted at angular frequencies of
    $\Omega_{\mathrm{r}}+\Omega_{\mathrm{p}}$ and $2\Omega_{\mathrm{r}}$
    \citep{Zimmermann:1980ba}. Previously, \citet{VanDenBroeck:2004wj}
    treated $\kappa$ as small as $\gamma^{2}$ and got a new line. It is at
    the second order of $\gamma$, but still at the first order of $\kappa$.
    The continuous GWs corresponding to this spectral line have an angular
    frequency of $2(\Omega_{\mathrm{r}}+\Omega_{\mathrm{p}})$. Here in our
    work, we treat $\kappa$ and $\gamma$ independently and expand the
    waveform to the second order that includes $\gamma^2$, $\kappa^2$, and
    $\gamma \kappa$. We find three new spectral lines at angular
    frequencies of $\Omega_{\mathrm{r}}-\Omega_{\mathrm{p}}$,
    $\Omega_{\mathrm{r}}+3\Omega_{\mathrm{p}}$ and
    $2(\Omega_{\mathrm{r}}-\Omega_{\mathrm{p}})$. We consider it a natural
    extension of the results in \citet{VanDenBroeck:2004wj}.
  \end{itemize}
%--

%---------------------------------------------------------------------
\section{Discussions}
\label{sec:disc}
%---------------------------------------------------------------------

The first detection of the coalescence of a binary NS system opened the avenue
for multimessenger astrophysics \citep{TheLIGOScientific:2017qsa,
GBM:2017lvd, Monitor:2017mdv}. In this paper, we discuss another
possibility to achieve multimessenger observation with electromagnetic and
GW detectors, namely the observation of precessing NSs. Multimessenger
astrophysics can be greatly advanced if a precessing NS is observed as a
pulsar via radio and/or X-ray telescopes, and in the mean time, its
continuous GW radiation is detected by the kilohertz laser-interferometric
GW detectors, including LIGO \citep{TheLIGOScientific:2014jea}, Virgo
\citep{TheVirgo:2014hva}, and KAGRA \citep{Akutsu:2018axf}. As we will see
below, this should surely provide invaluable constraints on the NS
structure, complementary to traditional observables, including masses,
radii and tidal deformabilities of NSs.

Radio/X-ray signals and continuous GWs from precessing triaxial NSs will
provide valuable information about the wobble angle, the nonaxisymmetry,
and the oblateness of the source. These measurements are ultimately related
to the long-standing question on the equation of state for supranuclear
matters inside NSs \citep{Lattimer:2000nx}. Below we take the case of a
small wobble angle as an example to discuss the extraction of physical
properties from such measurements \citep{VanDenBroeck:2004wj}.

For pulsar signals, the amplitude of the spin period residual in
Eq.~(\ref{eqn:delp}) at the frequencies of $\Omega_{\mathrm{p}}$ and $2\Omega_{\mathrm{p}}$
can be expressed as
%--
\begin{align}
    {\Delta P_{1}} =& { 1.6\times 10^{-10}} \cot\chi\left(
    \frac{P_{0}}{0.01\, \mathrm{s}} \right)^{2}
    \left(\frac{\Omega_{\mathrm{p}}}{10^{-5}\,\mathrm{rad\,s^{-1}}}\right)
    \left(\gamma +8\kappa \gamma\right) \,\mathrm{s} \,,\\
    {\Delta P_{2}} =& {8.0\times 10^{-11}} \left(1 + 2\cot^{2}\chi\right)
    \left( \frac{P_{0}}{0.01\,\mathrm{s}} \right)^{2} \left(
    \frac{\Omega_{\mathrm{p}}}{10^{-5}\, \mathrm{rad\,s^{-1}}}\right)
    \gamma^{2}\,\mathrm{s} \,,
\end{align}
%--
where $\Delta P_{1}$ is the amplitude of the spin period residual at the
frequency $\Omega_{\textrm{p}}$, and $\Delta P_{2}$ is the amplitude of the spin
period residual at the frequency $2\Omega_{\textrm{p}}$. 

The elliptic integral of the first kind $K(m)$ approaches $\pi/2$ in the
small-wobble-angle and small-nonaxisymmetry limit, which leads to
$\Omega_{\textrm{p}} \to \epsilon\, \Omega_{\textrm{r}}$. Therefore, in such a
limiting case, the precession angular frequency $\Omega_{\textrm{p}}$ can be
approximated as \citep{VanDenBroeck:2004wj}
%--
\begin{equation}
\label{eqn:get_ob}
    \Omega_{\mathrm{p}} \simeq \frac{\pi}{2 K(m)} \epsilon \, \Omega_{\mathrm{r}}\,.
\end{equation}
%--
The free precession period $T$ can be directly obtained from the positions
of spectral lines in the frequency domain of timing residuals. The wobble
angle $\gamma$, the nonaxisymmetry $\kappa$, and the magnetic inclination
angle $\chi$ cannot be fully determined with two spectral lines. But if the
nonaxisymmetry is small enough, the second-order contribution to the
amplitude of the first line can be ignored. Then the wobble angle $\gamma$
and the magnetic inclination angle $\chi$ can be determined. 

The pulse-width modulations will provide important information on the beam
shape of pulsars. In our work, we used a simple cone model
\citep{Gil:1984ads} to describe the modulations of pulse width. We find
that up to the second order, the pulse-width modulation is the same as that
in the biaxial case. From the perspective of observation, if the
pulse-width variations from precessing NSs are observed, the beam shape can
be inferred via different cuts by the line of sight \citep{Link:2001zr}.

From above discussions, we find that the inclusion of the nonaxisymmetry of
NS only slightly changes the timing residuals and the pulse width compared
with the biaxial results. The reason is that the parameter $m=16 \kappa
\gamma^{2}$ plays an important role in determining the behavior of free
precession. As $m$ approaches zero, the biaxial approximation is robust.
Even for a large nonaxisymmetry, if the wobble angle $\gamma \ll 1$, the
dynamics of the NS still only deviates from the biaxial one slightly. In
the case of large wobble angles and large nonaxisymmetries, the parameter
$m$ can be of order unity. Then the amplitudes of the harmonics are
correspondingly large in timing residuals. In this case, if the angle
between the beam and the line of sight changes during the free precession,
the observer might lose the radiating beam when the line of sight does not
cut the radiating region (see the upper panel in Fig.
\ref{fig.timing_width}). \Reply{Some pulsars display episodes of
interpulsing (i.e. a pulse occurring midway between successive main
pulses). The existence of interpulses in a specific system in principle
leads to a constraint on the orientation between the magnetic and rotation
axes \citep{Akgun:2005nd}, which may give us more information on the pulsar
geometries and pulse profiles. More exploration along this line is
worthwhile.}

For active pulsars, magnetospheric processes may affect the pulse signals
from precessing NSs and make the interpretation of free precession
complicated. For example, the precession may itself introduce changes on
the emission geometry and the activities of the magnetosphere
\citep{Link:2001zr}. Besides, the changes of the emission height can
contribute to pulse arrival time \citep{Link:2001zr}. Depending on the
properties of the observed pulsars, these complications need to be
considered. \Reply{Compared to the radio signals, the X-ray signals are
hardly affected by dispersion and scattering during the propagation. The
NICER mission can give phase-resolved X-ray spectroscopy for pulsars
\citep{Riley:2019yda, Bilous:2019knh}. High-precision X-ray timing for
millisecond pulsars have also been conducted \citep{Deneva:2019jal}. In the
near future, NICER X-ray timing of pulsars may help to track the
precession of NSs better.}

As a new observation window, GWs from precessing NSs can give complementary
physical information on these triaxial NSs. Following
\citet{VanDenBroeck:2004wj}, we present the procedures to extract physical
parameters from continuous GWs. We take the ``$\times$'' mode as an
example, and the discussion for the ``$+$'' mode is similar. For the
``$\times$''-polarized GW, the amplitudes of the first-order lines at
$\Omega_{\mathrm{r}}+\Omega_{\mathrm{p}}$ and $2\Omega_{\mathrm{r}}$ are
%--
\begin{align}
    {A_{\times}^{1}}= & {1.0\times 10^{-28}}\,\gamma \sin \iota
    \left(\frac{\epsilon}{4.9\times10^{-8}} \right) \left(
    \frac{f_{\mathrm{r}}}{100\,\mathrm{Hz}} \right)^{2}
    \left(\frac{10\,\mathrm{kpc}}{r}\right) \,,\\
    {A_{\times}^{2}}=& {3.3\times 10^{-27}}\,\kappa \cos\iota \left(
    \frac{\epsilon}{4.9\times10^{-8}} \right) \left(
    \frac{f_{\mathrm{r}}}{100\,\mathrm{Hz}} \right)^{2} \left(
    \frac{10\,\mathrm{kpc}}{r}\right) \,,
\end{align}
%--
where we have assumed that the moment of inertia of the NS is
$10^{45}\,\mathrm{g\,cm^{2}}$ and applied Eq.~(\ref{eqn:oblateness})
to estimate the oblateness at specific rotation frequency for a NS with
$M=1.4 \, M_{\odot}$ and $R=10 \,\textrm{km}$.

If the first-order lines are observed, the rotation frequency and the free
precession frequency can be determined. The inclination angle $\iota$ can
be obtained by comparing the amplitudes of different polarizations for the
two first-order lines. Note that the determination of the inclination angle
$\iota$ is model dependent with the radio signals \citep{Jones:2007zza}.
The derived $\iota$ from continuous GWs is less model dependent and can
help to probe the pulsar geometry \citep{Jones:2007zza}. In our work, the
inclination angle $\iota$ is needed to determine the pulse-width
modulations.

The oblateness, nonaxisymmetry, and wobble angle are degenerate in 
the first-order waveform. For the ``$\times$''-polarized GW, the amplitudes
of the second-order lines at $2(\Omega_{\mathrm{r}}+\Omega_{\mathrm{p}})$, 
$\Omega_{\mathrm{r}}-\Omega_{\mathrm{p}}$, $\Omega_{\mathrm{r}}+3\Omega_{\mathrm{p}}$, 
and $2(\Omega_{\mathrm{r}}-\Omega_{\mathrm{p}})$ are respectively 
%--
\begin{align}
    {A_{\times}^{3}}=& -{2.1\times 10^{-28}}\,\left(64\kappa^{2}+\gamma^{2}\right)  \nonumber\\
   & \times \cos \iota \left( \frac{\epsilon}{4.9\times10^{-8}} \right) \left(
   \frac{f_{\mathrm{r}}}{100\,\mathrm{Hz}}\right)^{2}\left(
   \frac{10\,\mathrm{kpc}}{r} \right)\,,
\end{align}
%--
\begin{align}
   {A_{\times}^{4}}=& {1.5\times 10^{-27}}\,\gamma \kappa \sin \iota \left(
   \frac{\epsilon}{4.9\times10^{-8}} \right) \left(
   \frac{f_{\mathrm{r}}}{100\,\mathrm{Hz}} \right)^{2}\left(
   \frac{10\,\mathrm{kpc}}{r}\right) \,,
\end{align}
%--
\begin{align}
    {A_{\times}^{5}}=& - {2.1\times 10^{-28}} \,\gamma \kappa\sin \iota
    \left( \frac{\epsilon}{4.9\times10^{-8}} \right) \left(
    \frac{f_{\mathrm{r}}}{100\,\mathrm{Hz}} \right)^{2}\left(
    \frac{10\,\mathrm{kpc}}{r}\right) \,,
\end{align}
%--
\begin{align}
    {A_{\times}^{6}}= & {1.3\times 10^{-26}} \,\kappa^{2} \cos\iota \left( \frac{\epsilon}{4.9\times10^{-8}} \right)
    \left( \frac{f_{\mathrm{r}}}{100\,\mathrm{Hz}} \right)^{2} \left(
    \frac{10\,\mathrm{kpc}}{r}\right) \,.
\end{align}
%--
Theoretically, by comparing the amplitudes of the two first-order lines and one of the
second-order lines, the wobble angle $\gamma$ and the nonaxisymmetry
$\kappa$ can be determined. Then, one can obtain the parameter
$m=16\kappa\gamma^{2}$ so that the oblateness can be determined using
Eq.~(\ref{eqn:get_ob}).

From the observational perspective, however, these amplitudes at the second order
are very small, and unlikely to be detectable with the Advanced LIGO/Virgo
detectors. Besides, if the coherent time of the observation is shorter than
the free precession period, the free precession angular frequency
$\Omega_{\textrm{p}}$ cannot be resolved in frequency domain. However, in the
optimistic situation when they are observed with the next-generation GW
detectors (e.g., the Einstein Telescope and Cosmic Explorer \citep{Hild:2010id,Sathyaprakash:2012jk,Punturo:2010zz,Evans:2016mbw}), 
they can be used to infer the oblateness, nonaxisymmetry, and wobble
angle of the star. The distance to the NS and the moment of inertia always enter the
waveform through the combination $I_{3}/r$. Therefore, we cannot obtain
them independently. By inserting an educated guess of $I_{3}$, the distance
to the NS can be roughly determined \citep{VanDenBroeck:2004wj}. Or
conversely, if the distance can be determined via parallax or dispersion
measure in pulsar timing data, one can get a measurement of $I_3$, thus
putting new constraints on the equation of state. Detailed analysis along
this line is beyond the scope of this paper, and we leave it to future
study.

%---------------------------------------------------------------------
\section{Summary}
\label{sec:sum}
%---------------------------------------------------------------------

To summarize, in this paper we describe both the analytical and numerical
methods to calculate the dynamical evolution of precessing triaxial rigid
bodies. We discuss the timing residuals and the pulse-width modulations for
precessing triaxial NSs. We also present concrete examples of the timing
residuals and the pulse-width modulations for large and small wobble
angles. For the GWs from triaxial precessing NSs, after reviewing the
general solution of the quadrupole waveform \citep{Zimmermann:1980ba} and
showing examples of the waveform in both time and frequency domains, we
extend the work by \citet{VanDenBroeck:2004wj} at the second order by
relaxing the assumption on the small parameters $\gamma$ and $\kappa$. We
obtain three new lines in the continuous GWs spectra, which might be useful 
for future continuous GW analysis using the third-generation ground-based 
detectors \citep{Evans:2016mbw}, depending on the distance of the sources. 
If the prospects of the multimessenger astrophysics
discussed in this work become reality, numerous information on the shape of
NSs and the equation of state of supranuclear matters will be obtained,
enabling a new frontier for fundamental physics.

%---------------------------------------------------------------------
\section*{Acknowledgements}
%---------------------------------------------------------------------
\Reply{We thank the anonymous referee for helpful comments and
suggestions.}
This work was supported by the National Natural Science Foundation of China
(11975027, 11991053, 11721303 and 11673002), the Young Elite Scientists
Sponsorship Program by the China Association for Science and Technology
(2018QNRC001), the Max Planck Partner Group Program funded by the Max Planck
Society, and the High-performance Computing Platform of Peking University. It
was partially supported by the Strategic Priority Research Program of the
Chinese Academy of Sciences through the Grant No. XDB23010200.  L. Sun is a
member of the LIGO Laboratory.  LIGO was constructed by the California
Institute of Technology and Massachusetts Institute of Technology with funding
from the United States National Science Foundation, and operates under
cooperative agreement PHY--1764464. Advanced LIGO was built under award
PHY--0823459. 

%---------------------------------------------------------------------
\section*{Data availability}
%---------------------------------------------------------------------

The data underlying this article will be shared on reasonable request to
the corresponding author.



\bibliographystyle{mnras}
\bibliography{refs} % if your bibtex file is called example.bib

%%%%%%%%%%%%%%%%%%%%%%%%%%%%%%%%%%%%%%%%%%%%%%%%%%

% Don't change these lines
\bsp	% typesetting comment
\label{lastpage}
\end{document}